# The Coming of Age of Nucleic Acid Vaccines during COVID-19

This manuscript ([permalink](#)) was automatically generated from [greenelab/covid19-review@3668a38](#) on January 24, 2023. It is also available as a [PDF](#). It represents one section of a larger evolving review on SARS-CoV-2 and COVID-19 available at https://greenelab.github.io/covid19-review/

**This in progress manuscript is not intended for the general public.** This is a review paper that is authored by scientists for an audience of scientists to discuss research that is in progress. If you are interested in guidelines on testing, therapies, or other issues related to your health, you should not use this document. Instead, you should collect information from your local health department, the [CDC's guidance](#), or your own government.

# Authors

- **Halie M. Rando** 0000-0001-7688-1770 rando2 tamefoxtime · Department of Systems Pharmacology and Translational Therapeutics, University of Pennsylvania, Philadelphia, Pennsylvania, United States of America; Department of Biochemistry and Molecular Genetics, University of Colorado Anschutz School of Medicine, Aurora, Colorado, United States of America; Center for Health AI, University of Colorado Anschutz School of Medicine, Aurora, Colorado, United States of America; Department of Biomedical Informatics, University of Colorado Anschutz School of Medicine, Aurora, Colorado, United States of America · Funded by the Gordon and Betty Moore Foundation (GBMF 4552); the National Human Genome Research Institute (R01 HG010067)

- **Ronan Lordan** 0000-0001-9668-3368 RLordan el_ronan · Institute for Translational Medicine and Therapeutics, Perelman School of Medicine, University of Pennsylvania, Philadelphia, PA 19104-5158, USA; Department of Medicine, Perelman School of Medicine, University of Pennsylvania, Philadelphia, PA 19104, USA; Department of Systems Pharmacology and Translational Therapeutics, Perelman School of Medicine, University of Pennsylvania; Philadelphia, PA 19104, USA

- **Likhitha Kolla** 0000-0002-1169-906X likhithakolla lkolla2018 · Perelman School of Medicine, University of Pennsylvania, Philadelphia, Pennsylvania, United States of America · Funded by NIH Medical Scientist Training Program T32 GM07170

- **Elizabeth Sell** 0000-0002-9658-1107 esell17 · Perelman School of Medicine, University of Pennsylvania, Philadelphia, Pennsylvania, United States of America

- **Alexandra J. Lee** 0000-0002-0208-3730 ajlee21 · Department of Systems Pharmacology and Translational Therapeutics, University of Pennsylvania, Philadelphia,

- Pennsylvania, United States of America · Funded by the Gordon and Betty Moore Foundation (GBMF 4552)

- **Nils Wellhausen** 0000-0001-8955-7582 nilswellhausen Department of Systems Pharmacology and Translational Therapeutics, University of Pennsylvania, Philadelphia, Pennsylvania, United States of America

- **Amruta Naik** 0000-0003-0673-2643 NAIKA86 Children's Hospital of Philadelphia, Philadelphia, PA, United States of America

- **Jeremy P. Kamil** 0000-0001-8422-7656 Department of Microbiology and Immunology, Louisiana State University Health Sciences Center Shreveport, Shreveport, Louisiana, USA

- **COVID-19 Review Consortium**

- **Anthony Gitter** 0000-0002-5324-9833 agitter anthonygitter Department of Biostatistics and Medical Informatics, University of Wisconsin-Madison, Madison, Wisconsin, United States of America; Morgridge Institute for Research, Madison, Wisconsin, United States of America · Funded by John W. and Jeanne M. Rowe Center for Research in Virology

- **Casey S. Greene** 0000-0001-8713-9213 cgreene GreeneScientist Department of Systems Pharmacology and Translational Therapeutics, University of Pennsylvania, Philadelphia, Pennsylvania, United States of America; Childhood Cancer Data Lab, Alex's Lemonade Stand Foundation, Philadelphia, Pennsylvania, United States of America; Department of Biochemistry and Molecular Genetics, University of Colorado Anschutz School of Medicine, Aurora, Colorado, United States of America; Center for Health AI, University of Colorado Anschutz School of Medicine, Aurora, Colorado, United States of America; Department of Biomedical Informatics, University of Colorado Anschutz School of Medicine, Aurora, Colorado, United States of America · Funded by the Gordon and Betty Moore Foundation (GBMF 4552); the National Human Genome Research Institute (R01 HG010067)

**COVID-19 Review Consortium:** Vikas Bansal, John P. Barton, Simina M. Boca, Joel D Boerckel, Christian Brueffer, James Brian Byrd, Stephen Capone, Shikta Das, Anna Ada Dattoli, John J. Dziak, Jeffrey M. Field, Soumita Ghosh, Anthony Gitter, Rishi Raj Goel, Casey S. Greene, Marouen Ben Guebila, Daniel S. Himmelstein, Fengling Hu, Nafisa M. Jadavji, Jeremy P. Kamil, Sergey Knyazev, Likhitha Kolla, Alexandra J. Lee, Ronan Lordan, Tiago Lubiana, Temitayo Lukan, Adam L. MacLean, David Mai, Serghei Mangul, David Manheim, Lucy D'Agostino McGowan, Jesse G. Meyer, Ariel I. Mundo, Amruta Naik, YoSon Park, Dimitri Perrin, Yanjun Qi, Diane N. Rafizadeh, Bharath Ramsundar, Halie M. Rando, Sandipan Ray, Michael P. Robson, Vincent Rubinetti, Elizabeth Sell, Lamonica Shinholster, Ashwin N. Skelly, Yuchen Sun, Yusha Sun, Gregory L Szeto, Ryan Velazquez, Jinhui Wang, Nils Wellhausen

Authors with similar contributions are ordered alphabetically.

# 1     Abstract

In the 21$^{st}$ century, several emergent viruses have posed a global threat. Each pathogen has emphasized the value of rapid and scalable vaccine development programs. The ongoing SARS-CoV-2 pandemic has made the importance of such efforts especially clear.

New biotechnological advances in vaccinology allow for recent advances that provide only the nucleic acid building blocks of an antigen, eliminating many safety concerns. During the COVID-19 pandemic, these DNA and RNA vaccines have facilitated the development and deployment of vaccines at an unprecedented pace. This success was attributable at least in part to broader shifts in scientific research relative to prior epidemics; the genome of SARS-CoV-2 was available as early as January 2020, facilitating global efforts in the development of DNA and RNA vaccines within two weeks of the international community becoming aware of the new viral threat. Additionally, these technologies that were previously only theoretical are not only safe but also highly efficacious.

Although historically a slow process, the rapid development of vaccines during the COVID-19 crisis reveals a major shift in vaccine technologies. Here, we provide historical context for the emergence of these paradigm-shifting vaccines. We describe several DNA and RNA vaccines and in terms of their efficacy, safety, and approval status. We also discuss patterns in worldwide distribution. The advances made since early 2020 provide an exceptional illustration of how rapidly vaccine development technology has advanced in the last two decades in particular and suggest a new era in vaccines against emerging pathogens.

# 2     Importance

The SARS-CoV-2 pandemic has caused untold damage globally, presenting unusual demands on but also unique opportunities for vaccine development. The development, production, and distribution of vaccines is imperative to saving lives, preventing severe illness, and reducing the economic and social burdens caused by the COVID-19 pandemic. Although vaccine technologies that provide the DNA or RNA sequence of an antigen had never previously been approved for use in humans, they have played a major role in the management of SARS-CoV-2. In this review we discuss the history of these vaccines and how they have been applied to SARS-CoV-2. Additionally, given that the evolution of new SARS-CoV-2 variants continues to present a significant challenge in 2022, these vaccines remain an important and evolving tool in the biomedical response to the pandemic.

# 3     Introduction

The SARS-CoV-2 virus emerged at the end of 2019 and soon spread around the world. In response, the Coalition for Epidemic Preparedness Innovations quickly began coordinating global health agencies and pharmaceutical companies to develop vaccines, as vaccination is one of the primary approaches available to combat the effects of a virus. Vaccines can bolster the immune response to a virus at both the individual and population levels, thereby reducing fatalities and severe illness and potentially driving a lower rate of infection even for a highly infectious virus like SARS-CoV-2. However, vaccines have historically required a lengthy development process due to both the experimental and regulatory demands.

As we review in a companion manuscript (1), vaccine technologies prior to the COVID-19 pandemic were largely based on triggering an immune response by introducing a virus or one of its components. Such vaccines are designed to induce an adaptive immune response without causing the associated viral illness. Each time a virus emerges that poses a significant global threat, as has happened several times over the past 20 years, the value of a rapid vaccine response is underscored. With progressive biotechnological developments, this objective has become increasingly tangible.

In the current century, significant advances in vaccine development have largely been built on genomics, as is somewhat unsurprising given the impact of the Genomic Revolution across all biology. This shift towards nucleic acid-based technologies opens a new frontier in vaccinology, where just the sequence encoding an antigen can be introduced to induce an immune response. While other platforms can carry some risks related to introducing all or part of a virus (1), nucleic acid-based platforms eliminate these risks entirely. Additionally, vaccine technologies that could be adjusted for novel viral threats are appealing because this modular approach would mean they could enter trials quickly in response to a new pathogen of concern.

# 4 Honing a 21$^{st}$ Century Response to Emergent Viral Threats

Recently, vaccine technologies have been developed and refined in response to several epidemics that did not reach the level of destruction caused by COVID-19. Emergent viral threats of the 21$^{st}$ century include severe acute respiratory syndrome (SARS), the H1N1 influenza strain known as swine flu, Middle East respiratory syndrome (MERS), Ebola virus disease, COVID-19, and, most recently, monkeypox, all of which have underscored the importance of a rapid global response to a new infectious virus. Because the vaccine development process has historically been slow, the use of vaccines to control most of these epidemics was limited.

One of the more successful recent vaccine development programs was for H1N1 influenza. This program benefited from the strong existing infrastructure for influenza vaccines along with the fact that regulatory agencies had determined that vaccines produced using egg- and cell-based platforms could be licensed under the regulations used for a strain change (2). Although a monovalent H1N1 vaccine was not available before the pandemic peaked in the United States of America (U.S.A.) and Europe, it became available soon afterward as a stand-alone vaccine that was eventually incorporated into commercially available seasonal influenza vaccines (2). Critiques of the production and distribution of the H1N1 vaccine have stressed the need for alternative development-and-manufacturing platforms that can be readily adapted to new pathogens.

Efforts to develop such approaches had been undertaken prior to the COVID-19 pandemic. DNA vaccine development efforts began for SARS-CoV-1 but did not proceed past animal testing (3). Likewise, the development of viral-vectored Ebola virus vaccines was undertaken, but the pace of vaccine development was behind the spread of the virus from early on (4). Although a candidate Ebola vaccine V920 showed promise in preclinical and clinical testing, it did not receive breakthrough therapy designation until the summer of 2016, by which time the Ebola outbreak was winding down (5). Therefore, the COVID-19 pandemic has been the first case where vaccines have been available early enough to significantly influence outcomes at the global scale.

The pandemic caused by SARS-CoV-2 has highlighted a confluence of circumstances that positioned vaccine development as a key player in efforts to control the virus and mitigate its damage. This virus did not follow the same trajectory as other emergent viruses of recent note, such as SARS-CoV-1, MERS-CoV, and Ebola virus, none of which presented a global threat for such a sustained duration (see visualization in (6)). Spread of the SARS-CoV-2 virus has remained out of control in many parts of the world into 2022, especially with the emergence of novel variants exhibiting increased rates of transmission (7). While, for a variety of reasons, SARS-CoV-2 was not controlled as rapidly as the viruses underlying prior 21$^{st}$ century epidemics, vaccine development technology had also progressed based on these and other prior viral threats to the point that a rapid international vaccine development response was possible.

# 5    Development of COVID-19 Vaccines using DNA and RNA Platforms

Vaccine development programs for COVID-19 emerged very quickly. The first administration of a dose of a COVID-19 vaccine to a human trial participant occurred on March 16, 2020 (8, 9), marking an extremely rapid response to the emergence of SARS-CoV-2. Within a few weeks of this first trial launching, at least 78 vaccine development programs were active (9), and by September 2020, there were over 180 vaccine candidates against SARS-CoV-2 in development (10). As of December 2, 2022, 50 SARS-CoV-2 vaccines have been approved world wide and 28 are being administered throughout the world, with 13.0 billion doses administered across 223 countries. The first critical step towards developing a vaccine against SARS-CoV-2 was characterizing the viral target, which happened extremely early in the COVID-19 outbreak with the sequencing and dissemination of the viral genome in early January 2020 (11) (Figure 1). This genomic information allowed for an early identification of the sequence of the Spike (S) protein (Figure 1), which is the antigen and induces an immune response (12, 13).

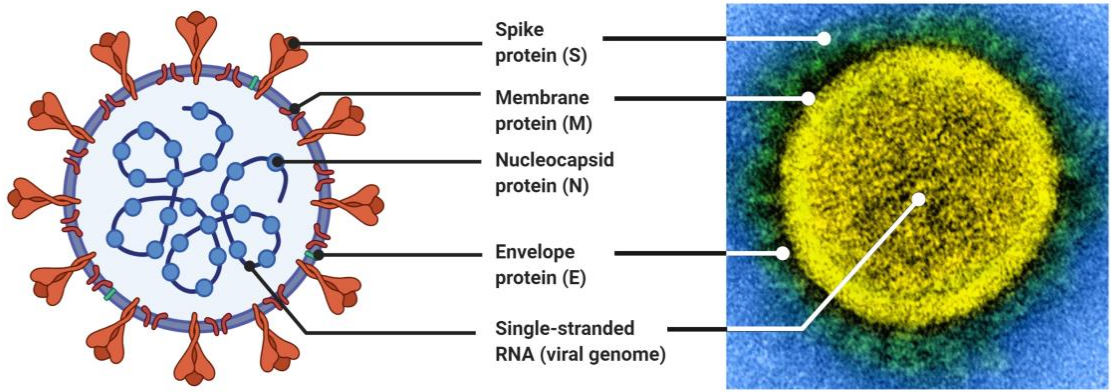

*Figure 1: **Structure of the SARS-CoV-2 virus.** The development of vaccines depends on the immune system recognizing the virus. Here, the structure of SARS-CoV-2 is represented both in the abstract and against a visualization of the virion. The abstracted visualization was made using BioRender ([14](#)) using the template "Human Coronavirus Structure" by BioRender (August 2020) ([15](#)). The microscopy was conducted by the National Institute of Allergy and Infectious Diseases ([16](#)).*

During the development process, one measure used to assess whether a vaccine candidate is likely to provide protection is serum neutralizing activity ([17](#)). This assay evaluates the presence of antibodies that can neutralize, or prevent infection by, the virus in question. Often, titration is used to determine the extent of neutralization activity. However, unlike in efforts to develop vaccines for prior viral threats, the duration of the COVID-19 pandemic has made it possible to also test vaccines in phase III trials where the effect of the vaccines on a cohort's likelihood of contracting SARS-CoV-2 was evaluated.

# 6 Theory and Implementation of Nucleic Acid Vaccines

Biomedical research in the 21$^{st}$ century has been significantly influenced by the genomic revolution. While traditional methods of vaccine development, such as inactivated whole viruses are still used today ([1](#)), vaccine development is no exception. The shift towards omics-based approaches to vaccine development began to take hold with the meningococcal type B vaccine, which was developed using reverse vaccinology in the early 2010s ([18](#), [19](#)). Under this approach, the genome of a pathogen is screened to identify potential vaccine targets ([19](#)), and pathogens of interest are then expressed *in vitro* and tested in animal models to determine their immunogenicity ([19](#)). In this way, the genomic revolution catalyzed a fundamental shift in the development of vaccines. Such technologies could revolutionize the role of vaccines given their potential to address one of the major limitations of vaccines today and facilitate the design of therapeutic, rather than just prophylactic, vaccines ([20](#)).

Nucleic-acid based approaches share an underlying principle: a vector that delivers the information needed to produce an antigen. When the host cells manufacture the antigen, it can then trigger an immune response. The fact that no part of the virus is introduced aside from the genetic code of the antigen means that these vaccines carry no risk of infection. Such approaches build on subunit vaccination strategies, where a component of a virus (e.g., an antigenic protein) is delivered by the vaccine. Platforms based on genomic sequencing began to be explored beginning in the 1980s as genetic research became increasingly feasible. Advances in genetic engineering allowed for gene sequences of specific viral antigens to be grown *in vitro* (21). Studies also demonstrated that model organisms could be induced to construct antigens that would trigger an immune response (22–24). These two developments sparked interest in whether it could be possible to identify any or all of the antigens encoded by a virus's genome and train the immune response to recognize them.

The delivery and presentation of antigens is fundamental to inducing immunity against a virus. Vaccines that deliver nucleic acids allow the introduction of foreign substances to the body to induce both humoral and cellular immune responses (25). Delivering a nucleic acid sequence to host cells allows the host to synthesize an antigen without exposure to a viral threat (25). Host-synthesized antigens can activate both humoral and cellular immunity (25), as they can be presented in complex with major histocompatibility complex (MHC) I and II, which can activate either T- or B-cells (25). In contrast, prior approaches activated only MHC II (24). Because these vaccines encode specific proteins, providing many of the benefits of a protein subunit vaccine, they do not carry any risk of DNA being live, replicating, or spreading, and their manufacturing process lends itself to scalability (25). Here, opportunities can be framed in terms of the central dogma of genetics: instead of directly providing the proteins from the infectious agents, vaccines developers are exploring the potential for the delivery of DNA or RNA to induce the cell to produce proteins from the virus that in turn induce a host immune response.

# 7 Cross-Platform Considerations in Vaccine Development

Certain design decisions are relevant to vaccine development across multiple platforms. One applies to the platforms that deliver the antigen, which in the case of SARS-CoV-2 vaccines is the S protein. The prefusion conformation of the S protein, which is the structure before the virus fuses to the host cell membrane, is metastable (26), and the release of energy during membrane fusion drives this process forward following destabilization (27, 28). Due to the significant conformational changes that occur during membrane fusion (29–31), S protein immunogens that are stabilized in the prefusion conformation are of particular interest, especially because a prefusion stabilized *Middle East respiratory syndrome-related coronavirus* (MERS-CoV) S antigen was found to elicit an improved antibody response (32). Moreover, the prefusion conformation offers an opportunity to target S2, a region of the S protein that accumulates mutations at a slower rate (32–34) (see also (7)). Vaccine developers can stabilize the prefusion conformer by selecting versions of the S protein containing mutations that lock the position (35). The immune response to the Spike protein when it is stabilized in this conformation is improved over other S structures (36). Thus, vaccines that use this prefusion stabilized conformation are expected to not only offer improved immunogenicity, but also be more resilient to the accumulation of mutations in SARS-CoV-2.

Another cross-platform consideration is the use of adjuvants. Adjuvants include a variety of molecules or larger microbial-related products that affect the immune system broadly or an immune response of interest. They can either be comprised of or contain immunostimulants or immunomodulators. Adjuvants are sometimes included within vaccines in order to enhance the immune response. Different adjuvants can regulate different types of immune responses, so the type or combination of adjuvants used in a vaccine will depend on both the type of vaccine and concern related to efficacy and safety. A variety of possible mechanisms for adjuvants have been investigated (37–39).

Due to viral evolution, vaccine developers are in an arms race with a pathogen that benefits from mutations that reduce its susceptibility to adaptive immunity. The evolution of several variants of concern (VOC) presents significant challenges for vaccines developed based on the index strain identified in Wuhan in late 2019. We discuss these variants in depth elsewhere in the COVID-19 Review Consortium project (40). To date, the most significant variants of concern identified are Alpha (2020), Beta (2020), Gamma (2020), Delta (2021), Omicron (2021), and related Omicron subvariants (2022). The effectiveness or efficacy (i.e., trial or real-world prevention, respectively) of vaccines in the context of these variants is discussed where information is available.

# 8 DNA Vaccine Platforms

DNA vaccine technologies have developed slowly over the past thirty years. These vaccines introduce a vector containing a DNA sequence that encodes antigen(s) selected to induce a specific immune response (24). Early attempts revealed issues with low immunogenicity (22, 24, 41). Additionally, initial skepticism about the approach suggested that DNA vaccines might bind to the host genome or induce autoimmune disease (25, 42), but pre-clinical and clinical studies have consistently disproved this hypothesis and indicated DNA vaccines to be safe (41). Another concern, antibiotic resistance introduced during the plasmid selection process, did remain a concern during this initial phase of development (25), but this issue was resolved through strategic vector design (43, 44). However, for many years, the immunogenicity of DNA vaccines failed to reach expectations (25). Several developments during the 2010s led to greater efficacy of DNA vaccines (25). However, no DNA vaccines had been approved for use in humans prior to the COVID-19 pandemic (41, 45). As of December 2, 2022, 10 vaccines have been approved worldwide (Table 1). These vaccines fall into two categories, vaccines that are vectored with a plasmid and those that are vectored with another virus.

*Table 1: DNA vaccines approved in at least one country (46) as of December 2, 2022.*

| Vaccine | Company | Platform |
| --- | --- | --- |
| iNCOVACC | Bharat Biotech | non replicating viral vector |
| Convidecia | CanSino | non replicating viral vector |
| Convidecia Air | CanSino | non replicating viral vector |

| Vaccine | Company | Platform |
| --- | --- | --- |
| Gam-COVID-Vac | Gamaleya | non replicating viral vector |
| Sputnik Light | Gamaleya | non replicating viral vector |
| Sputnik V | Gamaleya | non replicating viral vector |
| Jcovden | Janssen (Johnson & Johnson) | non replicating viral vector |
| Vaxzevria | Oxford/AstraZeneca | non replicating viral vector |
| Covishield (Oxford/ AstraZeneca formulation) | Serum Institute of India | non replicating viral vector |
| ZyCoV-D | Zydus Cadila | plasmid vectored |

## 8.1 Plasmid-Vectored DNA Vaccines

Many DNA vaccines use a plasmid vector-based approach, where the sequence encoding the antigen(s) against which an immune response is sought are cultivated in a plasmid and delivered directly to an appropriate tissue ([47](#)). Plasmids can also be designed to act as adjuvants by targeting essential regulators of pathways such as the inflammasome or simply just specific cytokines ([42](#), [48](#)). The DNA itself may also stimulate the innate immune response ([24](#), [44](#)). Once the plasmid brings the DNA sequence to an antigen-presenting cell (APC), the host machinery can be used to construct antigen(s) from the transported genetic material, and the body can then synthesize antibodies in response ([25](#)). The vectors are edited to remove extra sequences ([44](#)). These types of manufacturing advances have improved the safety and throughput of this platform ([44](#)).

### 8.1.1 Prior Applications

In the 1990s and 2000s, DNA vaccines delivered via plasmids sparked significant scientific interest, leading to a large number of preclinical trials ([25](#)). Early preclinical trials primarily focused on long-standing disease threats, including viral diseases such as rabies and parasitic diseases such as malaria, and promising results led to phase I testing of the application of this technology to human immunodeficiency virus (HIV), influenza, malaria, and other diseases of concern during this period ([25](#)). Although they were well-tolerated, these early attempts to develop vaccines were generally not very successful in inducing immunity to the target pathogen, with either limited T-cell or limited neutralizing antibody responses observed ([25](#)).

Early plasmid-vectored DNA vaccine trials targeted HIV and subsequently diseases of worldwide importance such as malaria and hepatitis B ([49](#)). The concern with these early development projects was immunogenicity, not safety ([49](#)). Around the turn of the millennium, a hepatitis B vaccine development program demonstrated that these vaccines can induce both antibody and cellular immune response ([50](#)). Prior to COVID-19, however, plasmid-vectored DNA vaccines had been approved for commercial use only in veterinary populations ([51](#)–[53](#)).

Between 2005 and 2006, several DNA vaccines were developed for non-human animal populations, including against viruses including a rhabdovirus in fish (54), porcine reproductive and respiratory syndrome virus (55), and West Nile virus in horses (56). Within the past five years, additional plasmid-vectored vaccines for immunization against viruses were developed against a herpesvirus (in mice) (57) and an alphavirus (in fish) (58).

### 8.1.2  Applications to COVID-19

Several plasmid-vectored DNA vaccines have been developed against COVID-19 (Table 1). In fact, the ZyCoV-D vaccines developed by India's Zydus Cadila is the first plasmid-vectored DNA vaccine to receive approval or to be used in human medicine (59–61). Another plasmid-vectored DNA vaccine, INO-4800 (62), was developed by Inovio Pharmaceuticals Technology that uses electroporation as an adjuvant. Electroporation was developed as a solution to the issue of limited immunogenicity by increasing the permeability of cell membranes by delivering electrical pulses (63). It has been shown that electroporation can enhance vaccine efficacy by up to 100-fold, as measured by increases in antigen-specific antibody titers (64). The temporary formation of pores through electroporation facilitates the successful transportation of macromolecules into cells, allowing cells to robustly take up INO-4800 for the production of an antibody response. For INO-4800, a plasmid-vectored vaccine is delivered through intradermal injection which is then followed by electroporation with a device known as CELLECTRA® (65). The safety of the CELLECTRA® device has been studied for over seven years, and these studies support the further development of electroporation as a safe vaccine delivery method (63).

These vaccines therefore represent implementations of a new platform technology. In particular, they offer the advantage of a temperature-stable vaccine, facilitating worldwide administration (66). Although an exciting development in DNA vaccines, the cost of vaccine manufacturing and electroporation may make scaling the use of this technology for prophylactic use for the general public difficult.

### 8.1.3  Trial Safety and Immunogenicity

The INO-4800 trials began with a phase I trial evaluating two different doses administered as a two-dose series (65). This trial found the vaccine to be safe, with only six adverse events (AEs) reported by 39 participants, all grade 1, and effective, with all but three participants of 38 developing serum IgG binding titers to the SARS-CoV-2 S protein (65). In the phase II trial of 401 adults at high risk of exposure to SARS-CoV-2 similarly supported the safety and efficacy of INO-4800. Only one treatment-related AE was observed and the vaccine was found to be associated with a significant increase in neutralizing activity (66). Results of phase III trials are not yet available (67–70).

Trials of ZyCoV-D have progressed further. This vaccine uses a plasmid to deliver the expression-competent Spike protein and IgE signal peptides to the vaccinee (71). During the phase I trial, vaccination with a needle versus a needle-free injection system was evaluated, and the vaccine can now be administered without a needle (59, 60). A phase III trial enrolling over 27,000 patients found no difference in AEs between the placebo and treatment groups and estimated the efficacy of ZyCoV-D to be 66.6% (72). It was authorized for people ages 12 and older (61) The highly portable design offers advantages over traditional vaccines (71), especially as the emergence of variants continues to challenge the effectiveness of vaccines. As of August 2022, ZyCoV-D has only been approved in India (73) and is not tracked by Our World in Data (74).

### 8.1.4 Real-World Safety and Effectiveness

In terms of the ability of plasmid-vectored vaccines to neutralize VOC, varying information is available. The situation for ZyCoV-D is somewhat different, as their phase III trial occurred during the Delta wave in India (72). At present, no major press releases have addressed the vaccine's ability to neutralize Omicron and related VOC, but reporting suggests that the manufacturers were optimistic about the vaccine in light of the Omicron variant as of late 2021 (75).

As for INO-4800, studies have examined whether the induced immune response can neutralize existing VOC. They assessed neutralization of several VOC relative to the index strain and found no difference in neutralization between the index strain and the Gamma VOC (P.1) (76). However, neutralization of the Alpha and Beta VOC was significantly lower (approximately two and seven times, respectively) (76). These findings are in line with the shifts in effectiveness reported for other vaccines (1). In addition to loss of neutralizing activity due to viral evolution, studies have also evaluated the decline in neutralizing antibodies (nAbs) induced by INO-4800 over time. Levels of nAbs remained statistically significant relative to the pre-vaccination baseline for six months (77). Administration of a booster dose induced a significant increase of titers relative to their pre-booster levels (77). Given the timing of this trial (enrollment between April and July 2020), it is unlikely that participants were exposed to VOC associated with decreased efficacy.

In light of the emergence of VOC against which many vaccines show lower effectiveness, Inovio Pharmaceuticals began to develop a new vaccine with the goal of improving robustness against known and future VOC (78). Known as INO-4802, this vaccine was designed to express a pan-Spike immunogen (79). Booster studies in rodents (80) and non-human primates (79) suggest that it may be more effective than INO-4800 in providing immunity to VOC such as Delta and Omicron when administered as part of a heterologous boost regimen, although boosting with INO-4800 was also very effective in increasing immunity in rhesus macaques (79). Therefore, boosting is likely to be an important strategy for this vaccine, especially as the virus continues to evolve.

## 8.2 Viral-Vectored DNA Vaccines

Plasmids are not the only vector that can be used to deliver sequences associated with viral antigens. Genetic material from the target virus can also be delivered using a second virus as a vector. Viral vectors have emerged as a safe and efficient method to furnish the nucleotide sequences of an antigen to the immune system (81). The genetic content of the vector virus is often altered to prevent it from replicating, but replication-competent viruses can also be used under certain circumstances (82). Once the plasmid or viral vector brings the DNA sequence to an APC, the host machinery can be used to construct antigen(s) from the transported genetic material, and the host can then synthesize antibodies in response ([25](#)).

One of the early viral vectors explored was adenovirus, with serotype 5 (Ad5) being particularly effective ([25](#)). This technology rose in popularity during the 2000s due to its being more immunogenic in humans and non-human primates than plasmid-vectored DNA vaccines ([25](#)). In the 2000s, interest also arose in utilizing simian adenoviruses as vectors because of the reduced risk that human vaccine recipients would have prior exposure resulting in adaptive immunity ([25](#), [83](#)), and chimpanzee adenoviruses were explored as a potential vector in the development of a vaccine against MERS-CoV ([84](#)).

Today, various viral-vector platforms including poxviruses ([85](#), [86](#)), adenoviruses ([87](#)), and vesicular stomatitis viruses ([88](#), [89](#)) are being developed, Viral-vector vaccines are able to induce both an antibody and cellular response; however, the response is limited due to the immunogenicity of the viral vector used ([87](#), [90](#)). An important consideration in identifying potential vectors is the immune response to the vector. Both the innate and adaptive immune responses can potentially respond to the vector, limiting the ability of the vaccine to transfer information to the immune system ([91](#)). Different vectors are associated with different levels of reactogenicity; for example, adenoviruses elicit a much stronger innate immune response than replication deficient adeno-associated viruses derived from parvoviruses ([91](#)). Additionally, using a virus circulating widely in human populations as a vector presents additional challenges because vaccine recipients may already have developed an immune response to the vector ([92](#)). Furthermore, repeated exposure to adenoviruses via viral-vectored DNA vaccines may increase reactivity to these vectors over time, presenting a challenge that will need to be considered in long-term development of these vaccines ([93](#), [94](#)).

### 8.2.1 Prior Applications

There are several viral vector vaccines that are available for veterinary use ([25](#), [95](#)), but prior to the COVID-19 pandemic, only one viral vector vaccine was approved by the United States' Food and Drug Administration (FDA) for use in humans. This vaccine is vectored with a recombinant vesicular stomatitis virus and targeted against the Ebola virus ([96](#)). Additionally, several phase I and phase II clinical trials for other vaccines are ongoing ([81](#)), and the technology is currently being explored for its potential against numerous infectious diseases including malaria ([97](#), [98](#)), Ebola ([99](#)–[101](#)), and HIV ([102](#), [103](#)).

The threat of MERS and SARS initiated interest in the application of viral vector vaccines to human coronaviruses ([84](#)), but efforts to apply this technology to these pathogens had not yet led to a successful vaccine candidate. In the mid-to-late 2000s, adenoviral vectored vaccines against SARS were found to induce SARS-CoV-specific IgA in the lungs of mice ([104](#)) but were later found to offer incomplete protection in ferret models ([105](#)). Gamaleya National Center of Epidemiology and Microbiology in Moscow sought to use an adenovirus platform for the development of vaccines for MERS-CoV and Ebola virus, although neither of the previous vaccines were internationally licensed ([106](#)).

In 2017, results were published from an initial investigation of two vaccine candidates against MERS-CoV containing the MERS-CoV *S* gene vectored with chimpanzee adenovirus, Oxford

University #1 (ChAdOx1), a replication-deficient chimpanzee adenovirus ([107]). This study reported that a candidate containing the complete S protein sequence induced a stronger neutralizing antibody response in mice than candidates vectored with modified vaccinia virus Ankara.

The candidate was pursued in additional research, and in the summer of 2020 results of two studies were published. The first reported that a single dose of ChAdOx1 MERS induced an immune response and inhibited viral replication in macaques ([108]). The second reported promising results from a phase I trial that administered the vaccine to adults and measured safety, tolerability, and immune response ([109]).

### 8.2.2 Application to COVID-19

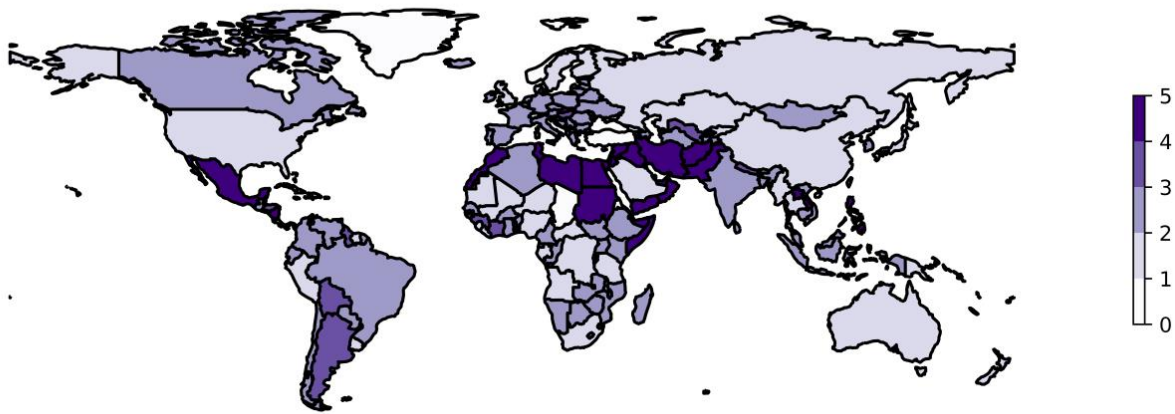

*Figure 2: **Worldwide availability of vaccines developed using non-replicating viral vectors.** This figure reflects the number of vaccines using non-replicating viral vectors that were available in each country as of January 22, 2023. These data are retrieved from Our World in Data ([74]) and plotted using geopandas ([110]). See https://greenelab.github.io/covid19-review/ for the most recent version of this figure, which is updated daily. Note that this figure draws from a different data source than Table [1] and does not necessarily include data for every vaccine developed within this category.*

While not all of the above results were available at the time that vaccine development programs against SARS-CoV-2 began, at least three viral vector vaccines have also been developed against SARS-CoV-2 (Figure [2]). First, a collaboration between AstraZeneca and researchers at the University of Oxford successfully applied a viral vector approach to the development of a vaccine against SARS-CoV-2 using the replication-deficient ChAdOx1 vector modified to encode the S protein of SARS-CoV-2 ([111]). In a phase I trial, the immunogenic potential of vaccine candidate ChAdOx1 nCoV-19 was demonstrated through the immune challenge of two animal models, mice and rhesus macaques ([111]). In a phase I/II trial, patients receiving the ChAdOx1 nCoV-19 vaccine developed antibodies to the SARS-CoV-2 Spike protein that peaked by day 28, with these levels remaining stable until a second observation at day 56 ([112]).

Second, a viral vector approach was applied by Russia's Gamaleya Research Institute of Epidemiology and Microbiology to develop Sputnik V, a replication-deficient recombinant adenovirus (rAd) vaccine that combines two adenovirus vectors, rAd26-S and rAd5-S, that express the full-length SARS-CoV-2 Spike glycoprotein. These vectors are intramuscularly administered individually using two separate vaccines in a prime-boost regimen. The rAd26-S is administered first, followed by rAd5-S 21 days later. Both vaccines deliver $10^{11}$ viral particles per dose. This approach is designed to overcome any potential pre-existing immunity to adenovirus in the population (113), as some individuals may possess immunity to Ad5 (114). Sputnik V is the only recombinant adenovirus vaccine to utilize two vectors.

Third, Janssen Pharmaceuticals, Inc., a subsidiary of Johnson & Johnson, developed a viral vector vaccine in collaboration with and funded by the United States' "Operation Warp Speed" (115, 116). The vaccine candidate JNJ-78436735, formerly known as Ad26.COV2-S, is a monovalent vaccine that is composed of a replication-deficient adenovirus serotype 26 (Ad26) vector expressing the stabilized prefusion S protein of SARS-CoV-2 (36, 117). Unlike the other two viral vector vaccines available to date, JNJ-78436735 requires only a single dose, a characteristic that was expected to aid in global deployment (118). JNJ-78436735 was selected from among a number of initial candidate designs (36) and tested *in vivo* in Syrian golden hamsters and Rhesus macaques to assess safety and immunogenicity (36, 118–120). The JNJ-78436735 candidate was selected for its favorable immunogenicity profile and ease of manufacturability (36, 118–120) and was found to confer protection against SARS-CoV-2 in macaques even after six months (121). The one- versus two-dose regimen was then tested in volunteers through a phase I/IIa trial (117, 122). A major difference between this vaccine and the other two in this category is that the S protein immunogen is stabilized in its prefusion conformation, while in the Sputnik V and AstraZeneca vaccines it is not.

As of January 22, 2023, data describing the distribution of 5 viral-vectored vaccines in 202 countries are available (Figure 2). ChAdOx1 nCoV-19 was first approved for emergency use on December 30, 2020 in the U.K. (123). Sputnik V was available soon after, and early as January 2021, Sputnik V had been administered to 1.5 million Russians (124) and began distributing doses to other countries within Europe such as Belarus, Bosnia-Herzegovina, Hungary, San Marino, Serbia, and Slovakia (125–127).

### 8.2.3  Trial Estimates of Safety and Efficacy

The first DNA viral-vectored vaccine for which efficacy estimates became available was AstraZeneca's ChAdOx1 nCoV-19. In December 2020, preliminary results of the phase III trial were released detailing randomized control trials conducted in the United Kingdom (U.K.), Brazil, and South Africa between April and November 2020 (12). These trials compared ChAdOx1 nCoV-19 to a control, but the design of each study varied; pooling data across studies indicated an overall efficacy of 70.4%. For Sputnik V, the phase III trial indicated an overall vaccine efficacy of 91.6% for symptomatic COVID-19 (128). As for Janssen, the vaccine was well-tolerated, and across all regions studied, it was found to be 66.9% effective after 28 days for the prevention of moderate to severe COVID-19 and to be 81.7% effective for the prevention of laboratory-confirmed severe COVID-19 (129). There were no COVID-19-associated deaths in the vaccine group. However, the emergence of the Beta variant in the South African trial population was associated with a slightly reduced efficacy (64% two weeks after receipt), and all of the COVID-19-associated deaths in the trial occurred in the South African placebo cohort (129). In February 2021, the FDA issued an EUA for the Janssen vaccine based on interim results from the phase III trial (130, 131).

Two of the three vaccines have faced a number of criticisms surrounding the implementation of their clinical trials. In the race to develop vaccines against SARS-CoV-2, President Vladimir Putin of Russia announced the approval of the Sputnik V vaccines on August 11, 2020 in the absence of clinical evidence (132). A press release on November 11, 2020 indicated positive results from an interim analysis of the phase III Sputnik V trials, which reported 92% efficacy in 16,000 volunteers (133). However, this release came only two days after both Pfizer and BioNTech reported that their vaccines had an efficacy over 90%, which led to significant skepticism of the Russian findings for a myriad of reasons including the lack of a published protocol and the "reckless" approval of the vaccine in Russia months prior to the publication of the interim results of the phase III trial (133, 134). Consequently, many international scientific agencies and public health bodies expressed concern that due diligence to the clinical trial process was subverted for the sake of expediency, leading many to question the safety and efficacy of Sputnik V (132, 135, 136). Despite regulatory, safety, and efficacy concerns, pre-orders for 1 billion doses of the Sputnik V were reported within days of the vaccine's approval in Russia (132). Almost a month later, the phase I/II trial data was published (137) It wasn't until February 2021, six months after its approval in Russia, that interim results of the phase III trial were released (128). This publication reported a VE of 91% and a low rate of serious AEs, although there were several serious AEs that were determined not to be associated with the vaccine by an independent data monitoring committee about which little other information was released (138).

AstraZeneca's clinical trial also faced criticism. The trial was paused in September 2020 following a severe adverse event in one participant (139). It was restarted soon after (140), but it seems that the recent pause was not mentioned to the FDA during a call the morning before the story broke (141). Additionally, individual sites within the trial employed somewhat different designs but were combined for analysis. For example, in South Africa, the trial was double-blinded, whereas in the U.K. and Brazil it was single-blinded, and one of the two trials carried out in the U.K. evaluated two dosing regimens (low dose or standard dose, both followed by standard dose). Some of the trials used a meningococcal conjugate vaccine (MenACWY) as a control, while others used saline. Data was pooled across countries for analysis, a design decision that was approved by regulators but raised some questions when higher efficacy was reported in a subgroup of patients who received a low-dose followed by a standard dose. This group came about because some participants in the U.K. were erroneously primed with a much lower dose, which turned out to have higher efficacy than the intended dose (142). Combining the data then led to confusion surrounding the VE, as VE varied widely among conditions (e.g., 62% VE in the standard dose group vs 90% in the group that received a low prime dose (12)). Subsequent research, however, suggests that reducing the prime dose may, in fact, elicit a superior immune response in the long-term despite a lower initial response (143). Therefore, this error may serendipitously improve efficacy of vaccine-vectored vaccines broadly.

### 8.2.4 Real-World Safety and Efficacy

Following the trials, additional concerns have been raised about some of these vaccines. Within a few days to a few weeks following their first dose of the AstraZeneca vaccine, three women developed extensive venous sinus thrombosis (144). In March 2021, administration of the vaccine was paused in several European countries while a possible link to thrombotic events was investigated (145), as these adverse events had not been observed in clinical trials, but the European Medicine Agency (EMA) soon determined that 25 events were not related to the vaccine (146). The following month, the United States paused administration of the Janssen vaccine for ten days due to 15 similar AEs (147, 148), but the EMA, U.S. Centers for Disease Control, and the FDA's Advisory Committee on Immunization Practices again identified the events as being very rare and the benefits of the vaccine as likely to outweigh its risks (149–152). In Denmark and Norway, population-based estimates suggested AstraZeneca's vaccine increased incidence of venous thromboembolic events by 11 cases over baseline per 100,000 doses (153). Estimates of the incidence in other western countries have also been low (154). In the US, thromboembolic events following the Janssen vaccine have also been very rare (150). Subsequently, a potential mechanism was identified: the adenovirus vector binding to platelet factor 4 (155, 156). Because this adverse event is so rare, the risk is likely still outweighed by the risks associated with contracting COVID-19 (157), which is also associated with thrombotic events) (148, 158). Similarly, concerns about Guillain-Barré syndrome arose in connection to the Janssen vaccine, but these events have similarly been determined to be very rare and the benefits to outweigh the risks (152).

Given that vaccines from multiple platforms are now widely available, people at increased risk of a specific severe AE may have options to pursue vaccination with a platform that does not carry such risks. For example, a woman in the U.S. with a history of thromboembolic concerns might feel more comfortable with an mRNA vaccine (described below), where such AEs have not been identified in association with COVID-19 vaccination. However, within the U.S.A., no clear framework has been established for advising patients on whether a specific vaccine may be preferable for their individual concerns now that vaccines based on three different technologies are widely available (see (1) for information about Novavax, which is a protein subunit vaccine).

## 9 mRNA Vaccines

Building on DNA vaccine technology, RNA vaccines are an even more recent advancement for vaccine development. Interest in messenger RNA (mRNA) vaccines emerged around 1990 following *in vitro* and animal model studies that demonstrated that mRNA could be transferred into cells (159, 160). mRNA contains the minimum information needed to create a protein (160). RNA vaccines are therefore nucleic-acid based modalities that code for viral antigens against which the human body elicits a humoral and cellular immune response.

The strategy behind mRNA vaccines operates one level above the DNA: instead of directly furnishing the gene sequence associated with an antigen to the host, it provides the mRNA transcribed from the DNA sequence. The mRNA is transcribed *in vitro* and delivered to cells via lipid nanoparticles (LNP) (161). It is recognized by ribosomes *in vivo* and then translated and modified into functional proteins (162). The resulting intracellular viral proteins are displayed on surface MHC proteins, provoking a strong CD8+ T cell response as well as a CD4$^+$ T cell and B cell-associated antibody responses (162). mRNA is naturally not very stable and can degrade quickly in the extracellular environment or the cytoplasm. The LNP covering protects the mRNA from enzymatic degradation outside of the cell (163). Codon optimization to prevent secondary structure formation and modifications of the poly-A tail as well as the 5' untranslated region to promote ribosomal complex binding can increase mRNA expression in cells. Furthermore, purifying out double-stranded RNA and immature RNA with fast performance liquid chromatography and high performance liquid chromatography technology will improve translation of the mRNA in the cell (162, 164).

There are three types of RNA vaccines: non-replicating, *in vivo* self-replicating, and *in vitro* dendritic cell non-replicating (165). Non-replicating mRNA vaccines consist of a simple open reading frame for the viral antigen flanked by the 5' UTR and 3' poly-A tail. *In vivo* self-replicating vaccines encode a modified viral genome derived from single-stranded, positive sense RNA alphaviruses (162, 164). The RNA genome encodes the viral antigen along with proteins of the genome replication machinery, including an RNA polymerase. Structural proteins required for viral assembly are not included in the engineered genome (162). Self-replicating vaccines produce more viral antigens over a longer period of time, thereby evoking a more robust immune response (165). Finally, *in vitro* dendritic cell non-replicating RNA vaccines limit transfection to dendritic cells. Dendritic cells are potent antigen-presenting immune cells that easily take up mRNA and present fragments of the translated peptide on their MHC proteins, which can then interact with T cell receptors. Ultimately, primed T follicular helper cells can stimulate germinal center B cells that also present the viral antigen to produce antibodies against the virus (166). These cells are isolated from the patient, then grown and transfected *ex vivo* (167). They can then be reintroduced to the patient (167).

In addition to the benefits of nucleic acid vaccines broadly, mRNA confers specific advantages compared to DNA vaccines and other platforms (168). Some of these advantages fall within the domain of safety. Unlike DNA vaccines, mRNA technologies are naturally degradable and non-integrating, and they do not need to cross the nuclear membrane in addition to the plasma membrane for their effects to be seen (162). Additionally, the half life can be regulated by the contents of the 5' and 3' untranslated regions (169). In comparison to vaccines that use live attenuated viruses, mRNA vaccines are non-infectious and can be synthetically produced in an egg-free, cell-free environment, thereby reducing the risk of a detrimental immune response in the host (170). Furthermore, mRNA vaccines are easily, affordably, and rapidly scalable, despite the fact that it took time to reach the scale needed to manufacture vaccines at a scale sufficient for the global population (168).

### 9.0.1 Prior Applications

Although mRNA vaccines have been developed for therapeutic and prophylactic purposes, none have previously been licensed or commercialized. Challenges were caused by the instability of mRNA molecules, the design requirements of an efficient delivery system, and the potential for mRNA to elicit either a very strong immune response or to stimulate the immune system in secondary ways (20, 171). As of the 2010s, mRNA was still considered a promising technology for future advances in vaccine development (160), but prior to 2020, no mRNA vaccines had been approved for use in humans, despite significant advances in the development of this technology (167). This approach showed promise in animal models and preliminary clinical trials for several indications, including rabies, coronavirus, influenza, and cytomegalovirus (172). Preclinical data previously identified effective antibody generation against full-length purified influenza hemagglutinin stalk-encoding mRNA in mice, rabbits, and ferrets (173). Similar immunological responses for mRNA vaccines were observed in humans in phase I and II clinical trials operated by the pharmaceutical-development companies Curevac and Moderna for rabies, flu, and zika (164). Positively charged bilayer LNPs carrying the mRNA attract negatively charged cell membranes, endocytose into the cytoplasm (163), and facilitate endosomal escape. LNPs can be coated with modalities recognized and engulfed by specific cell types, and LNPs that are 150 nm or less effectively enter into lymphatic vessels (163, 174). Therefore, while these technologies elegantly capitalize on decades of research in vaccine development as well as the tools of the genomic revolution, it was largely unknown prior to the SARS-CoV-2 pandemic whether this potential could be realized in a real-world vaccination effort.

### 9.0.2 Application to COVID-19

Table 2: mRNA vaccines approved in at least one country (46) as of December 2, 2022. As a note, this table includes licensing of existing mRNA technology, i.e., TAK-919 is used to describe Takeda's manufacturing of Moderna's formulation.

| Vaccine | Company |
| --- | --- |
| GEMCOVAC-19 | Gennova Biopharmaceuticals Limited |
| Spikevax | Moderna |
| Spikevax Bivalent Original/Omicron BA.1 | Moderna |
| Spikevax Bivalent Original/Omicron BA.4/BA.5 | Moderna |
| Comirnaty | Pfizer/BioNTech |
| Comirnaty Bivalent Original/Omicron BA.1 | Pfizer/BioNTech |
| Comirnaty Bivalent Original/Omicron BA.4/BA.5 | Pfizer/BioNTech |
| TAK-919 (Moderna formulation) | Takeda |
| AWcorna | Walvax |

Given the potential for mRNA technology to be quickly adapted for a new pathogen, it was favored as a potential vaccine against COVID-19, and fortunately, the prior work in mRNA vaccine development paid off, with 9 mRNA vaccines available in at least one country as of December 2, 2022 (Table 2). In the vaccines developed under this approach, the mRNA coding for a stabilized prefusion Spike protein, which is immunogenic (175), is furnished to the immune system in order to train its response.

Two vaccine candidates in this category emerged with promising phase III results at the end of 2020. Both require two doses approximately one month apart. The first was Pfizer/BioNTech's BNT162b2, which contains the full prefusion stabilized, membrane-anchored SARS-CoV-2 Spike protein in a vaccine formulation based on modified mRNA (modRNA) technology (176, 177). The second mRNA vaccine, mRNA-1273 developed by ModernaTX, is comprised by a conventional LNP-encapsulated RNA encoding a full-length prefusion stabilized S protein for

SARS-CoV-2 (178). The vaccine candidates developed against SARS-CoV-2 using mRNA vectors utilize similar principles and technologies, although there are slight differences in implementation among candidates such as the formulation of the platform and the specific components of the Spike protein encapsulated (e.g., the full Spike protein vs. the RBD alone) (179). As of January 22, 2023, 2 mRNA vaccines are available in 169 countries (Figure 3).

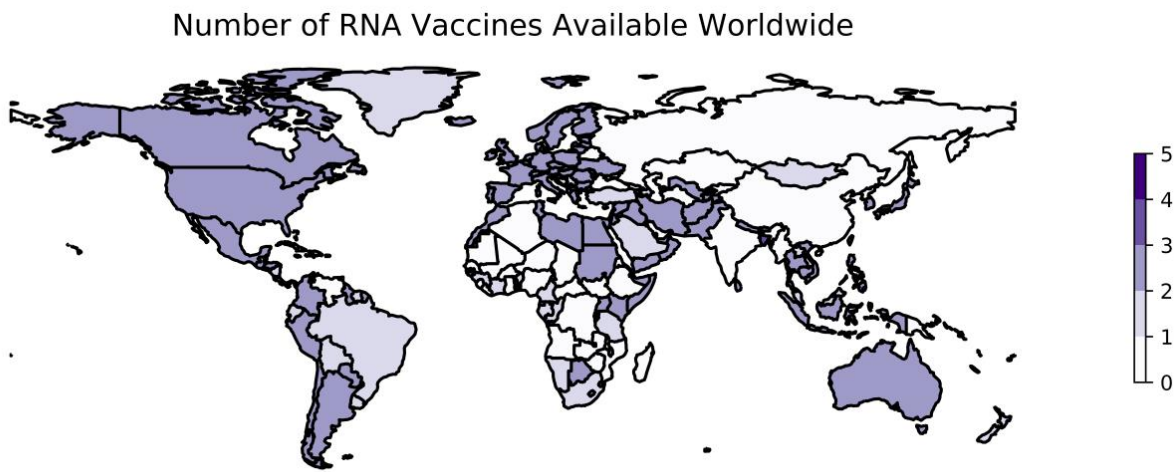

*Figure 3: **Worldwide availability of vaccines developed using mRNA.** This figure reflects the number of vaccines based on mRNA technology that were available in each country as of January 22, 2023. These data are retrieved from Our World in Data (74) and plotted using geopandas (110). See https://greenelab.github.io/covid19-review/ for the most recent version of this figure, which is updated daily. Note that this figure draws from a different data source than Table 2 and does not necessarily include data for every vaccine developed within this category.*

The rapid and simultaneous development of these vaccines was met with some controversy related to intellectual property (IP). First, the National Institutes of Health (NIH) and Moderna became involved in a patent dispute, after researchers at the NIH argued they were unfairly excluded from some patents filed based on their IP after they generated the stabilized modRNA sequence used in the vaccine (180). Ultimately, in late 2021, Moderna backed down on the patent application (181). However, in August 2022, the company filed their own suit against Pfizer/BioNTech over IP related to the modRNA used in the latter's COVID-19 vaccine (181, 182). The outcome of this suit remains to be seen.

### 9.0.3 Trial Safety and Immunogenicity

The VEs revealed by the Pfizer/BioNTech and Moderna clinical trials exceeded expectations. In a phase II/III multinational trial, the Pfizer/BioNTech's BNT162b2 vaccine was associated with a 95% efficacy against laboratory-confirmed COVID-19 and with mild-to-moderate local and systemic effects but a low risk of serious AEs when the prime-boost doses were administered 21 days apart (183). The ModernaTX mRNA-1273 vaccine was the second mRNA vaccine to release phase III results, despite being the first mRNA vaccine to enter phase I clinical trials and publish interim results of their phase III trial a few months later. Their study reported a 94.5% vaccine efficacy in preventing symptomatic COVID-19 in adults who received the vaccine at 99 sites around the United States (184). Similar to BNT162b2, the mRNA-1273 vaccine was associated with mild-to-moderate AEs but with a low risk of serious AEs (184). In late 2020, both vaccines received approval from the FDA under an emergency use authorization (185, 186), and these vaccines have been widely distributed, primarily in North America and the European Union (187). As the first mRNA vaccines to make it to market, these two highly efficacious vaccines demonstrate the power of this emerging technology, which has previously attracted scientific interest because of its potential to be used to treat non-infectious as well as infectious diseases.

### 9.0.4 Real-World Safety and Effectiveness

As vaccines were rolled out, one study sought to monitor their effectiveness in a real-world setting. Between December 2020 and April 2021, this prospective cohort study obtained weekly nasal swabs from 3,975 individuals at high risk of SARS-CoV-2 exposure (health care workers, frontline workers, etc.) within the United States (188). Among these participants, 3,179 (80%) had received at least one dose of an mRNA vaccine, and of those, 2,686 (84%) were fully vaccinated, corresponding to 68% of trial participants overall. For each vaccinated participant (defined here as having received at least dose 1 more than 7 days ago) whose sample tested positive for SARS-CoV-2, they categorized the viral lineage(s) present in the sample as well as in samples from 3-4 unvaccinated individuals matched by site and testing date. Overall efficacy of mRNA vaccines was estimated at 91% with full vaccination, similar to the reports from the clinical trials. The occurrence of fevers was also lower in individuals who were partially or fully vaccinated, and the duration of symptoms was approximately 6 days shorter. Among the five cases in fully vaccinated and 11 cases in partially vaccinated participants, the rate of infection by VOC was much higher than in the unvaccinated population (30% versus 10%), suggesting that the vaccine was less effective against the VOC than the index strain.

The WHO continues to monitor the emergence of variants and their impact on vaccine efficacy (189). In general, mRNA vaccines remain highly effective against severe illness and death, but the effectiveness against infection generally has declined. A study monitoring infections in a Minnesota cohort from January to July 2021 estimated that the effectiveness of the Moderna vaccine fell to 86% and Pfizer to 76%, although protection against hospitalization remained at 91% and 85%, respectively (190). In July of that year, as the Delta variant became dominant in the U.S.A., these estimates all fell, to an effectiveness of 76% for Moderna and 42% for Pfizer and effectiveness against hospitalization of 81% and 75%, respectively (190).

With the emergence of the Omicron VOC, vaccine effectiveness has likely further declined. A study in a diverse cohort in Southern California, U.S.A. found the effectiveness of the Moderna vaccine in participants who had received only the primary course to be 44% (191). A study in South Africa compared case and hospitalization records from a 4-week period where Omicron was dominant to a 2-month period where Delta was dominant and found that the effectiveness against hospitalization during the Omicron wave was approximately 70% compared to 93% during the Delta wave (192). Similarly, a large study in England of 2.5 million individuals suggested that not only the variants circulating, but also the time since vaccination, played a large role in vaccine effectiveness (193). Shortly after the BNT162b2 primary course, effectiveness against the Omicron VOC was as high as 65.5%, but this declined to below 10% by six months after the second dose. For mRNA-1273, the decline was from 75.1% to 14.9%.

Therefore, it is unsurprising that in spite of vaccination programs, infection rates and hospitalization rates climbed in early 2022 in many Western countries including the United States (194, 195), especially given that many places simultaneously began to loosen public health restrictions designed to reduce viral spread.

On the side of safety, the only major concern that has been raised is a possible link between mRNA vaccination and myocarditis, especially in young men (152). This concern began with case reports of a small number of cases of myocarditis following vaccination in several countries (196, 197). Following these reports, the Israeli Ministry of Health began surveillance to monitor the occurrence of myocarditis (198). They identified 283 cases, almost exactly half of which occurred following vaccination with Pfizer's BNT162b2. Close analysis of these cases determined that the vaccine did have a significant effect on the incidence of myocarditis; however, the rate of myocarditis remained low overall (198). The identification of young men as a population at particular risk of this AE was supported, and the risk was found to be greater after the second dose than the first. Both this study and a study evaluating data collected from US population-based surveillance identified an increased risk with additional doses (199). However, most findings suggest that this AE does not have long-term negative effects; a 2021 meta-analysis identified 69 cases, all of which resulted in full recovery (200). Although these events are very rare, as with the possible thromboembolic AEs associated with viral-vectored DNA vaccines, these findings suggest that it may be prudent to offer a framework for decision making for patients particularly concerned about specific AEs in settings where multiple vaccines are available.

# 10 Booster Doses

Due to waning effectiveness of vaccines over time, especially in light of viral evolution, boosters have emerged as an important strategy in retaining the benefits of vaccination over time. Booster shots are now recommended in many places, and boosters that account for multiple variants and strains of SARS-CoV-2 are now available in some places (201). For example, in the U.S.A., the FDA recently recommended bivalent booster doses designed to account for the Omicron VOC (202–204). In this case, bivalent refers to the fact that doses deliver both the original formulation and an updated vaccine designed for the Omicron subvariants circulating in summer 2022. The fact that the FDA did not require additional clinical trials from manufacturers for Omicron subvariants BA.4 and BA.5 specifically suggests that the rapid authorization of strain changes in response to emerging VOC may be increasingly attainable (205). Results suggest that this fourth dose offered at least a short-term increase in VE against Omicron subvariants and also provided additional protection against hospitalization (206).

Homologous booster doses have been investigated for most vaccines. For example, over 14,000 adults were administered a booster (second) dose of the Janssen Ad26.COV2.S vaccine (207). The booster dose was highly efficacious, with severe COVID-19 and hospitalization prevented almost completely in the vaccinated group. A booster dose was also found to improve immune response for Sputnik V vaccinees (208). For the AstraZeneca vaccine, a different approach was taken. In the interest of distributing first doses as widely as possible, in some places the time between the first and second doses was extended. One study assessed the immunogenicity and reactogenicity associated with delaying the second dose in the prime-boost series until up to 45 weeks after the first, reporting that an extended inter-dose period was associated with increased antibody titers 28 days after the second dose (209). This analysis also revealed that a third dose provided an additional boost in neutralizing activity (209).

Third and fourth doses have been introduced for at least some populations in many places in response to the Omicron variant. An early study in Israeli healthcare workers showed that the additional immunization was safe and immunogenic with antibody titers restored to peak-third dose titers. No severe illness was reported in the cohort studied (274 versus 426 age-matched controls), and vaccine efficacy against infection was reported at 30% for BNT162b2 and 11% for mRNA-1273 (210). Other studies reported that a third dose of BNT162b2 raised vaccine effectiveness to 67.2% for approximately the first month but that the effectiveness dropped to 45.7% (193). Reduced and even low efficacy against infection does not undermine the value of vaccination, considering the vaccines are intended to prevent severe disease, hospitalization, and death rather than infection generally. However, these findings do suggest that boosters will likely be needed as the virus continues to evolve.

Many trials have also investigated heterologous boosting approaches. In particular, the mRNA vaccines are a popular choice for booster doses regardless of primary series. In general, such approaches have been found to confer favorable immunogenicity relative to homologous boosters (e.g., (211–217) and many other studies). Due to remaining concerns about rare thromboembolic events, vaccinees who received AstraZeneca for their primary course are advised in some countries to seek a heterologous booster (218), although such guidances are not supported by the evidence, which indicates that the first dose of AstraZeneca is most likely to be linked to these rare events (219). In general, heterologous boosting with mRNA vaccines elicits a strong immune response. For patients who received BNT162b2 as a heterologous booster following a ChAdOx1 primary series, the vaccine effectiveness was estimated to be 62.4% initially, dropping to 39.6% after 10 weeks (193). For a heterologous mRNA-1273 booster, the effectiveness was estimated to be slightly higher (70.1% and 60.9% following ChAdOx1 and 73.9% to 64.4% following BNT162b2) (193). Therefore, subsequent booster doses may remain an ongoing component of strategies to combat SARS-CoV-2.

Although the vaccines developed based on the index strain remain highly effective at preventing severe illness and death, they serve much less utility at preventing illness broadly than they did early in the pandemic. Therefore, many manufacturers are exploring potential reformulations based on VOC that have emerged since the beginning of the pandemic. In June 2022, Moderna released data describing the effect of their bivalent mRNA booster, mRNA-1273.214, designed to protect against the Omicron variant (220). A 50 µg dose of mRNA-1273.214 was administered to 437 participants. One month later, the neutralizing geometric mean titer ratio was assessed against several variants of SARS-CoV-2, including Omicron. The immune response was higher against all variants assessed, including Omicron, than for boosting with the original formulation (mRNA-1273). Another formulation, mRNA-1273.211, developed based on the Beta variant, has been associated with durable protection as long as six months after dosing. The associated publications suggest that this novel formulation offers significant protection against Omicron and other VOC (221, 222). In August 2022, Pfizer also announced successful development of a new formulation effective against Omicron (223).

Modularity has been proposed as one of the advantages to developing DNA and mRNA vaccines. This design would allow for faster adaptation to viral evolution. However, in the arms race against SARS-CoV-2, the vaccines are still lagging behind the virus. This disadvantage may change as regulators become more familiar with these vaccines and as a critical mass of data is accumulated. Given the apparent need for boosters, interest has also emerged in whether updated formulations of SARS-CoV-2 vaccines can be administered along with annual flu vaccines to improve immunity to novel variants.

## 11  Conclusions

COVID-19 has seen the coming-of-age of vaccine technologies that have been in development since the late 20[th] century but had never before been authorized for use. Vaccines that employ DNA and RNA eliminate all concerns about potential infection due to the vaccine components. The vaccines described above demonstrate the potential for these technologies to facilitate a quick response to an emerging pathogen. Additionally, their efficacy in trials far exceeded expectations, especially in the case of RNA vaccines. These technologies hold significant potential to drive improvements in human health over the coming years.

Traditional vaccine technologies were built on the principle of using either a weakened version of the virus or a fragment of the virus. COVID-19 has highlighted the fact that in recent years, the field has undergone a paradigm shift towards reverse vaccinology. Reverse vaccinology emphasizes a discovery-driven approach to vaccine development based on knowledge of the viral genome ([224](#)). This strategy was explored during development of a DNA vaccine against the Zika virus ([225](#)). Though the disease was controlled before the vaccine became available ([2](#)), the response demonstrated the potential for modular technologies to facilitate a response to emerging viral threats ([225](#)). The potential for such vaccines to benefit the field of oncology has encouraged vaccine developers to invest in next-generation approaches, which has spurred the diversification of vaccine development programs ([25](#), [226](#)). As a result, during the COVID-19 pandemic, these modular technologies have taken center stage in controlling a viral threat for the first time.

The safety and efficacy of vaccines that use these new technologies has exceeded expectations. While there were rare reports of severe AEs such as myocarditis (mRNA platforms) and thromboembolic events (viral-vectored DNA platforms), widespread availability of both types of vaccines would allow individuals to choose (particularly relevant in this case because myocarditis has primarily been reported in men and thromboembolic events primarily in women). Estimates of efficacy have varied widely, but in all cases are high. Estimates of the efficacy of DNA vaccine platforms have typically fallen either in the range of approximately 67% (ZyCoV-D and Janssen) or 90% (Sputnik V). AstraZeneca's trial produced estimates in both ranges, with the standard dosage producing an efficacy of 62% and the lower prime dose producing a VE of 90%. The mRNA vaccine trials were somewhat higher, with VE estimated at approximately 95% for both the Moderna and Pfizer/BioNTech clinical trials. However, in all cases, the efficacy against severe illness and death were very high. Therefore, all of these vaccines are useful tools for combating COVID-19.

Furthermore, the fact that vaccine efficacy is not a static value has become particularly salient, as real-world effectiveness has changed with location and over time. COVID-19 vaccines have been challenged by the emergence of VOC. These VOC generally carry genetic mutations that code for an altered Spike protein (i.e., the antigen), so the antibodies resulting from immunization with vaccines developed from the index strain neutralize them less effectively ([227](#), [228](#)). Despite some reports of varying and reduced effectiveness or efficacy of the mRNA vaccines against the Alpha (B.1.1.7), Beta (B.1.351), and Delta (B.1.617.2) variants versus the original SARS-CoV-2 strain or the D614G variant ([229](#)–[231](#)), the greatest concern to date has been the Omicron variant (B.1.1.529), which was first identified in November 2021 ([228](#), [232](#)). As of March 2022, the Omicron variant accounted for 95% of all infections sequenced in the United States ([233](#)) and was linked to an increased risk of SARS-CoV-2 reinfection ([227](#)) and further infection of those who have been vaccinated with the mRNA vaccines ([234](#)).

One of the downsides of this leap in vaccine technologies, however, is that they have largely been developed by wealthy countries, including countries in the European Union, the United States, the U.K., and Russia. As a result, they are also largely available to residents of wealthy countries, primarily in Europe and North America. Although the VE of DNA vaccines tends to be lower than that of mRNA vaccines ([235](#)), they still provide excellent protection against severe illness and are much easier to distribute due to less complex demands for storage. Efforts such as COVAX that aim to expand access to vaccines developed by wealthy countries have not been as successful as hoped ([236](#)). Fortunately, vaccine development programs using more established technologies have been undertaken in many middle-income countries, and those vaccines have been more accessible globally ([1](#)). Additionally, efforts to develop new formulations of DNA vaccines in lower- and middle-income countries are increasingly being undertaken ([237](#)).

The modular nature of nucleic acid-based vaccine platforms has opened a new frontier in responding to emerging viral illnesses. The RNA vaccines received an EUA in only a few months more than it took to identify the pathogen causing SARS in 2002. Given the variety of options available for preventing severe illness and death, it is possible that certain vaccines may be preferable for certain demographics (e.g., young women might choose an mRNA vaccine to entirely mitigate the very low risk of blood clots ([152](#))). However, this option is likely only available to people in high-income countries. In lower-income countries, access to vaccines broadly is a more critical issue. Different vaccines may confer advantages in different countries, and vaccine development in a variety of cultural contexts is therefore important ([238](#)). Without widespread access to vaccines on the global scale, SARS-CoV-2 will continue evolving, presenting a threat to all nations.

# 1 Additional Items

## 1.1 Competing Interests

| Author | Competing Interests | Last Reviewed |
|---|---|---|
| Halie M. Rando | None | 2021-01-20 |

| Author | Competing Interests | Last Reviewed |
|---|---|---|
| Ronan Lordan | None | 2020-11-03 |
| Likhitha Kolla | None | 2020-11-16 |
| Elizabeth Sell | None | 2020-11-11 |
| Alexandra J. Lee | None | 2020-11-09 |
| Nils Wellhausen | None | 2020-11-03 |
| Amruta Naik | None | 2021-04-05 |
| Jeremy P. Kamil | None | 2022-10-11 |
| COVID-19 Review Consortium | None | 2021-01-16 |
| Anthony Gitter | Inventor of patent US-11410440-B2 assigned to the Wisconsin Alumni Research Foundation related to classifying activated T cells. | 2023-01-11 |
| Casey S. Greene | None | 2021-01-20 |

## 1.2 Author Contributions

| Author | Contributions |
|---|---|
| Halie M. Rando | Project Administration, Software, Visualization, Writing - Original Draft, Writing - Review & Editing |
| Ronan Lordan | Project Administration, Visualization, Writing - Original Draft, Writing - Review & Editing |
| Likhitha Kolla | Writing - Original Draft, Writing - Review & Editing |
| Elizabeth Sell | Writing - Review & Editing |
| Alexandra J. Lee | Writing - Review & Editing |
| Nils Wellhausen | Project Administration |
| Amruta Naik | Writing - Review & Editing |
| Jeremy P. Kamil | Writing - Review & Editing |
| COVID-19 Review Consortium | Project Administration |
| Anthony Gitter | Project Administration, Software, Writing - Review & Editing |
| Casey S. Greene | Conceptualization, Supervision, Writing - Original Draft |

## 1.3 Acknowledgements

We thank Nick DeVito for assistance with the Evidence-Based Medicine Data Lab COVID-19 TrialsTracker data. We thank Yael Evelyn Marshall who contributed writing (original draft) as well as reviewing and editing of pieces of the text but who did not formally approve the manuscript, as well as Ronnie Russell, who contributed text to and helped develop the structure of the manuscript early in the writing process and Matthias Fax who helped with writing and editing text related to diagnostics. We are also very grateful to James Fraser for suggestions about successes and limitations in the area of computational screening for drug repurposing. We are grateful to the following contributors for reviewing pieces of the text: Nadia Danilova, James Eberwine and Ipsita Krishnan.

## 1.4 References

1.      Rando HM, Lordan R, Lee AJ, Naik A, Wellhausen N, Sell E, Kolla L, Consortium CR, Gitter A, Greene CS. 2022. Application of Traditional Vaccine Development Strategies to SARS-CoV-2. 2208.08907arXiv. arXiv.

2.      Lurie N, Saville M, Hatchett R, Halton J. 2020. Developing Covid-19 Vaccines at Pandemic Speed. New England Journal of Medicine 382:1969–1973.

3.      Graham RL, Donaldson EF, Baric RS. 2013. A decade after SARS: strategies for controlling emerging coronaviruses. Nat Rev Microbiol 11:836–848.

4.      Cohen J. 2014. Ebola vaccine: Little and late. Science 345:1441–1442.

5.      Coller B-AG, Blue J, Das R, Dubey S, Finelli L, Gupta S, Helmond F, Grant-Klein RJ, Liu K, Simon J, Troth S, VanRheenen S, Waterbury J, Wivel A, Wolf J, Heppner DG, Kemp T, Nichols R, Monath TP. 2017. Clinical development of a recombinant Ebola vaccine in the midst of an unprecedented epidemic. Vaccine 35:4465–4469.

6.      Rando HM, Wellhausen N, Ghosh S, Lee AJ, Dattoli AA, Hu F, Byrd JB, Rafizadeh DN, Lordan R, Qi Y, Sun Y, Brueffer C, Field JM, Ben Guebila M, Jadavji NM, Skelly AN, Ramsundar B, Wang J, Goel RR, Park Y, COVID-19 Review Consortium Vikas Bansal, John P. Barton, Simina M. Boca, Joel D. Boerckel, Christian Brueffer, James Brian Byrd, Stephen Capone, Shikta Das, Anna Ada Dattoli, John J. Dziak, Jeffrey M. Field, Soumita Ghosh, Anthony Gitter, Rishi Raj Goel, Casey S. Greene, Marouen Ben Guebila, Daniel S. Himmelstein, Fengling Hu, Nafisa M. Jadavji, Jeremy P. Kamil, Sergey Knyazev, Likhitha Kolla, Alexandra J. Lee, Ronan Lordan, Tiago Lubiana, Temitayo Lukan, Adam L. MacLean, David Mai, Serghei Mangul, David Manheim, Lucy D'Agostino McGowan, Amruta Naik, YoSon Park, Dimitri Perrin, Yanjun Qi, Diane N. Rafizadeh, Bharath Ramsundar, Halie M. Rando, Sandipan Ray, Michael P. Robson, Vincent Rubinetti, Elizabeth Sell, Lamonica Shinholster, Ashwin N. Skelly, Yuchen Sun, Yusha Sun, Gregory L. Szeto, Ryan Velazquez, Jinhui Wang, Nils Wellhausen, Boca SM, Gitter A, Greene CS. 2021. Identification and Development of Therapeutics for COVID-19. mSystems 6:e0023321.

7.      Rando HM, MacLean AL, Lee AJ, Lordan R, Ray S, Bansal V, Skelly AN, Sell E, Dziak JJ, Shinholster L, D'Agostino McGowan L, Ben Guebila M, Wellhausen N, Knyazev S, Boca SM, Capone S, Qi Y, Park Y, Mai D, Sun Y, Boerckel JD, Brueffer C, Byrd JB, Kamil JP, Wang J, Velazquez R, Szeto GL, Barton JP, Goel RR, Mangul S, Lubiana T, COVID-19 Review Consortium Vikas Bansal, John P. Barton, Simina M. Boca, Joel D. Boerckel, Christian Brueffer, James Brian Byrd, Stephen Capone, Shikta Das, Anna Ada Dattoli, John J. Dziak, Jeffrey M. Field, Soumita Ghosh, Anthony Gitter, Rishi Raj Goel, Casey S. Greene, Marouen Ben Guebila, Daniel S. Himmelstein, Fengling Hu, Nafisa M. Jadavji, Jeremy P. Kamil, Sergey Knyazev,

Likhitha Kolla, Alexandra J. Lee, Ronan Lordan, Tiago Lubiana, Temitayo Lukan, Adam L. MacLean, David Mai, Serghei Mangul, David Manheim, Lucy D'Agostino McGowan, Amruta Naik, YoSon Park, Dimitri Perrin, Yanjun Qi, Diane N. Rafizadeh, Bharath Ramsundar, Halie M. Rando, Sandipan Ray, Michael P. Robson, Vincent Rubinetti, Elizabeth Sell, Lamonica Shinholster, Ashwin N. Skelly, Yuchen Sun, Yusha Sun, Gregory L. Szeto, Ryan Velazquez, Jinhui Wang, Nils Wellhausen, Gitter A, Greene CS. 2021. Pathogenesis, Symptomatology, and Transmission of SARS-CoV-2 through Analysis of Viral Genomics and Structure. mSystems 6:e0009521.

8.  Our Story. Moderna. https://www.modernatx.com/en-US/about-us/our-story?slug=about-us%2Four-story. Retrieved 5 December 2022.

9.  Thanh Le T, Andreadakis Z, Kumar A, Gómez Román R, Tollefsen S, Saville M, Mayhew S. 2020. The COVID-19 vaccine development landscape. Nat Rev Drug Discov 19:305–306.

10. Krammer F. 2020. SARS-CoV-2 vaccines in development. Nature 586:516–527.

11. Novel Coronavirus – China. https://www.who.int/emergencies/disease-outbreak-news/item/2020-DON233. Retrieved 5 December 2022.

12. Voysey M, Clemens SAC, Madhi SA, Weckx LY, Folegatti PM, Aley PK, Angus B, Baillie VL, Barnabas SL, Bhorat QE, Bibi S, Briner C, Cicconi P, Collins AM, Colin-Jones R, Cutland CL, Darton TC, Dheda K, Duncan CJA, Emary KRW, Ewer KJ, Fairlie L, Faust SN, Feng S, Ferreira DM, Finn A, Goodman AL, Green CM, Green CA, Heath PT, Hill C, Hill H, Hirsch I, Hodgson SHC, Izu A, Jackson S, Jenkin D, Joe CCD, Kerridge S, Koen A, Kwatra G, Lazarus R, Lawrie AM, Lelliott A, Libri V, Lillie PJ, Mallory R, Mendes AVA, Milan EP, Minassian AM, McGregor A, Morrison H, Mujadidi YF, Nana A, O'Reilly PJ, Padayachee SD, Pittella A, Plested E, Pollock KM, Ramasamy MN, Rhead S, Schwarzbold AV, Singh N, Smith A, Song R, Snape MD, Sprinz E, Sutherland RK, Tarrant R, Thomson EC, Török ME, Toshner M, Turner DPJ, Vekemans J, Villafana TL, Watson MEE, Williams CJ, Douglas AD, Hill AVS, Lambe T, Gilbert SC, Pollard AJ, Aban M, Abayomi F, Abeyskera K, Aboagye J, Adam M, Adams K, Adamson J, Adelaja YA, Adewetan G, Adlou S, Ahmed K, Akhalwaya Y, Akhalwaya S, Alcock A, Ali A, Allen ER, Allen L, Almeida TCDSC, Alves MPS, Amorim F, Andritsou F, Anslow R, Appleby M, Arbe-Barnes EH, Ariaans MP, Arns B, Arruda L, Azi P, Azi L, Babbage G, Bailey C, Baker KF, Baker M, Baker N, Baker P, Baldwin L, Baleanu I, Bandeira D, Bara A, Barbosa MAS, Barker D, Barlow GD, Barnes E, Barr AS, Barrett JR, Barrett J, Bates L, Batten A, Beadon K, Beales E, Beckley R, Belij-Rammerstorfer S, Bell J, Bellamy D, Bellei N, Belton S, Berg A, Bermejo L, Berrie E, Berry L, Berzenyi D, Beveridge A, Bewley KR, Bexhell H, Bhikha S, Bhorat AE, Bhorat ZE, Bijker E, Birch G, Birch S, Bird A, Bird O, Bisnauthsing K, Bittaye M, Blackstone K, Blackwell L, Bletchly H, Blundell CL, Blundell SR, Bodalia P, Boettger BC, Bolam E, Boland E, Bormans D, Borthwick N, Bowring F, Boyd A, Bradley P, Brenner T, Brown P, Brown C, Brown-O'Sullivan C, Bruce S, Brunt E, Buchan R, Budd W, Bulbulia YA, Bull M, Burbage J, Burhan H, Burn A, Buttigieg KR, Byard N, Cabera Puig I, Calderon G, Calvert A, Camara S, Cao M, Cappuccini F, Cardoso JR, Carr M, Carroll MW, Carson-Stevens A, Carvalho Y de M, Carvalho JAM, Casey HR, Cashen P, Castro T, Castro LC, Cathie K, Cavey A, Cerbino-Neto J, Chadwick J, Chapman D, Charlton S, Chelysheva I, Chester O, Chita S, Cho J-S, Cifuentes L, Clark E,

Clark M, Clarke A, Clutterbuck EA, Collins SLK, Conlon CP, Connarty S, Coombes N, Cooper C, Cooper R, Cornelissen L, Corrah T, Cosgrove C, Cox T, Crocker WEM, Crosbie S, Cullen L, Cullen D, Cunha DRMF, Cunningham C, Cuthbertson FC, Da Guarda SNF, da Silva LP, Damratoski BE, Danos Z, Dantas MTDC, Darroch P, Datoo MS, Datta C, Davids M, Davies SL, Davies H, Davis E, Davis J, Davis J, De Nobrega MMD, De Oliveira Kalid LM, Dearlove D, Demissie T, Desai A, Di Marco S, Di Maso C, Dinelli MIS, Dinesh T, Docksey C, Dold C, Dong T, Donnellan FR, Dos Santos T, dos Santos TG, Dos Santos EP, Douglas N, Downing C, Drake J, Drake-Brockman R, Driver K, Drury R, Dunachie SJ, Durham BS, Dutra L, Easom NJW, van Eck S, Edwards M, Edwards NJ, El Muhanna OM, Elias SC, Elmore M, English M, Esmail A, Essack YM, Farmer E, Farooq M, Farrar M, Farrugia L, Faulkner B, Fedosyuk S, Felle S, Feng S, Ferreira Da Silva C, Field S, Fisher R, Flaxman A, Fletcher J, Fofie H, Fok H, Ford KJ, Fowler J, Fraiman PHA, Francis E, Franco MM, Frater J, Freire MSM, Fry SH, Fudge S, Furze J, Fuskova M, Galian-Rubio P, Galiza E, Garlant H, Gavrila M, Geddes A, Gibbons KA, Gilbride C, Gill H, Glynn S, Godwin K, Gokani K, Goldoni UC, Goncalves M, Gonzalez IGS, Goodwin J, Goondiwala A, Gordon-Quayle K, Gorini G, Grab J, Gracie L, Greenland M, Greenwood N, Greffrath J, Groenewald MM, Grossi L, Gupta G, Hackett M, Hallis B, Hamaluba M, Hamilton E, Hamlyn J, Hammersley D, Hanrath AT, Hanumunthadu B, Harris SA, Harris C, Harris T, Harrison TD, Harrison D, Hart TC, Hartnell B, Hassan S, Haughney J, Hawkins S, Hay J, Head I, Henry J, Hermosin Herrera M, Hettle DB, Hill J, Hodges G, Horne E, Hou MM, Houlihan C, Howe E, Howell N, Humphreys J, Humphries HE, Hurley K, Huson C, Hyder-Wright A, Hyams C, Ikram S, Ishwarbhai A, Ivan M, Iveson P, Iyer V, Jackson F, De Jager J, Jaumdally S, Jeffers H, Jesudason N, Jones B, Jones K, Jones E, Jones C, Jorge MR, Jose A, Joshi A, Júnior EAMS, Kadziola J, Kailath R, Kana F, Karampatsas K, Kasanyinga M, Keen J, Kelly EJ, Kelly DM, Kelly D, Kelly S, Kerr D, Kfouri R de Á, Khan L, Khozoee B, Kidd S, Killen A, Kinch J, Kinch P, King LDW, King TB, Kingham L, Klenerman P, Knapper F, Knight JC, Knott D, Koleva S, Lang M, Lang G, Larkworthy CW, Larwood JPJ, Law R, Lazarus EM, Leach A, Lees EA, Lemm N-M, Lessa A, Leung S, Li Y, Lias AM, Liatsikos K, Linder A, Lipworth S, Liu S, Liu X, Lloyd A, Lloyd S, Loew L, Lopez Ramon R, Lora L, Lowthorpe V, Luz K, MacDonald JC, MacGregor G, Madhavan M, Mainwaring DO, Makambwa E, Makinson R, Malahleha M, Malamatsho R, Mallett G, Mansatta K, Maoko T, Mapetla K, Marchevsky NG, Marinou S, Marlow E, Marques GN, Marriott P, Marshall RP, Marshall JL, Martins FJ, Masenya M, Masilela M, Masters SK, Mathew M, Matlebjane H, Matshidiso K, Mazur O, Mazzella A, McCaughan H, McEwan J, McGlashan J, McInroy L, McIntyre Z, McLenaghan D, McRobert N, McSwiggan S, Megson C, Mehdipour S, Meijs W, Mendonça RNÁ, Mentzer AJ, Mirtorabi N, Mitton C, Mnyakeni S, Moghaddas F, Molapo K, Moloi M, Moore M, Moraes-Pinto MI, Moran M, Morey E, Morgans R, Morris S, Morris S, Morris HC, Morselli F, Morshead G, Morter R, Mottal L, Moultrie A, Moya N, Mpelembue M, Msomi S, Mugodi Y, Mukhopadhyay E, Muller J, Munro A, Munro C, Murphy S, Mweu P, Myasaki CH, Naik G, Naker K, Nastouli E, Nazir A, Ndlovu B, Neffa F, Njenga C, Noal H, Noé A, Novaes G, Nugent FL, Nunes G, O'Brien K, O'Connor D, Odam M, Oelofse S, Oguti B, Olchawski V, Oldfield NJ, Oliveira MG, Oliveira C, Oosthuizen A, O'Reilly P, Osborne P, Owen DRJ, Owen L, Owens D, Owino N, Pacurar M, Paiva BVB, Palhares EMF, Palmer S, Parkinson S, Parracho HMRT, Parsons K, Patel D, Patel B, Patel F, Patel K, Patrick-Smith M, Payne RO, Peng Y, Penn EJ, Pennington A, Peralta Alvarez MP, Perring J, Perry N, Perumal R, Petkar S, Philip T, Phillips DJ, Phillips J, Phohu MK, Pickup L, Pieterse S, Piper J, Pipini D, Plank M, Du Plessis J, Pollard S, Pooley J, Pooran A, Poulton I, Powers C, Presa FB, Price DA, Price V, Primeira M, Proud PC, Provstgaard-Morys S, Pueschel S, Pulido D, Quaid S, Rabara R,

Radford A, Radia K, Rajapaska D, Rajeswaran T, Ramos ASF, Ramos Lopez F, Rampling T, Rand J, Ratcliffe H, Rawlinson T, Rea D, Rees B, Reiné J, Resuello-Dauti M, Reyes Pabon E, Ribiero CM, Ricamara M, Richter A, Ritchie N, Ritchie AJ, Robbins AJ, Roberts H, Robinson RE, Robinson H, Rocchetti TT, Rocha BP, Roche S, Rollier C, Rose L, Ross Russell AL, Rossouw L, Royal S, Rudiansyah I, Ruiz S, Saich S, Sala C, Sale J, Salman AM, Salvador N, Salvador S, Sampaio M, Samson AD, Sanchez-Gonzalez A, Sanders H, Sanders K, Santos E, Santos Guerra MFS, Satti I, Saunders JE, Saunders C, Sayed A, Schim van der Loeff I, Schmid AB, Schofield E, Screaton G, Seddiqi S, Segireddy RR, Senger R, Serrano S, Shah R, Shaik I, Sharpe HE, Sharrocks K, Shaw R, Shea A, Shepherd A, Shepherd JG, Shiham F, Sidhom E, Silk SE, da Silva Moraes AC, Silva-Junior G, Silva-Reyes L, Silveira AD, Silveira MBV, Sinha J, Skelly DT, Smith DC, Smith N, Smith HE, Smith DJ, Smith CC, Soares A, Soares T, Solórzano C, Sorio GL, Sorley K, Sosa-Rodriguez T, Souza CMCDL, Souza BSDF, Souza AR, Spencer AJ, Spina F, Spoors L, Stafford L, Stamford I, Starinskij I, Stein R, Steven J, Stockdale L, Stockwell LV, Strickland LH, Stuart AC, Sturdy A, Sutton N, Szigeti A, Tahiri-Alaoui A, Tanner R, Taoushanis C, Tarr AW, Taylor K, Taylor U, Taylor IJ, Taylor J, te Water Naude R, Themistocleous Y, Themistocleous A, Thomas M, Thomas K, Thomas TM, Thombrayil A, Thompson F, Thompson A, Thompson K, Thompson A, Thomson J, Thornton-Jones V, Tighe PJ, Tinoco LA, Tiongson G, Tladinyane B, Tomasicchio M, Tomic A, Tonks S, Towner J, Tran N, Tree J, Trillana G, Trinham C, Trivett R, Truby A, Tsheko BL, Turabi A, Turner R, Turner C, Ulaszewska M, Underwood BR, Varughese R, Verbart D, Verheul M, Vichos I, Vieira T, Waddington CS, Walker L, Wallis E, Wand M, Warbick D, Wardell T, Warimwe G, Warren SC, Watkins B, Watson E, Webb S, Webb-Bridges A, Webster A, Welch J, Wells J, West A, White C, White R, Williams P, Williams RL, Winslow R, Woodyer M, Worth AT, Wright D, Wroblewska M, Yao A, Zimmer R, Zizi D, Zuidewind P. 2021. Safety and efficacy of the ChAdOx1 nCoV-19 vaccine (AZD1222) against SARS-CoV-2: an interim analysis of four randomised controlled trials in Brazil, South Africa, and the UK. The Lancet 397:99–111.

13. Duan L, Zheng Q, Zhang H, Niu Y, Lou Y, Wang H. 2020. The SARS-CoV-2 Spike Glycoprotein Biosynthesis, Structure, Function, and Antigenicity: Implications for the Design of Spike-Based Vaccine Immunogens. Front Immunol 11.

14. BioRender. https://biorender.com/. Retrieved 5 December 2022.

15. Human Coronavirus Structure. https://app.biorender.com/biorender-templates/figures/all/t-5f21e90283765600b08fbe9d-human-coronavirus-structure. Retrieved 5 December 2022.

16. New Images of Novel Coronavirus SARS-CoV-2 Now Available | NIH: National Institute of Allergy and Infectious Diseases. https://www.niaid.nih.gov/news-events/novel-coronavirus-sarscov2-images. Retrieved 5 December 2022.

17. Jackson LA, Anderson EJ, Rouphael NG, Roberts PC, Makhene M, Coler RN, McCullough MP, Chappell JD, Denison MR, Stevens LJ, Pruijssers AJ, McDermott A, Flach B, Doria-Rose NA, Corbett KS, Morabito KM, O'Dell S, Schmidt SD, Swanson PA, Padilla M, Mascola JR, Neuzil KM, Bennett H, Sun W, Peters E, Makowski M, Albert J, Cross K, Buchanan W, Pikaart-Tautges R, Ledgerwood JE, Graham BS, Beigel JH. 2020. An mRNA Vaccine against SARS-CoV-2 — Preliminary Report. New England Journal of Medicine 383:1920–1931.

18. Rhee JH. 2014. Towards Vaccine 3.0: new era opened in vaccine research and industry. Clin Exp Vaccine Res 3:1.

19. Seib KL, Zhao X, Rappuoli R. 2012. Developing vaccines in the era of genomics: a decade of reverse vaccinology. Clinical Microbiology and Infection 18:109–116.

20. Liu. 2019. A Comparison of Plasmid DNA and mRNA as Vaccine Technologies. Vaccines 7:37.

21. Plotkin S. 2014. History of vaccination. Proc Natl Acad Sci USA 111:12283–12287.

22. Cui Z. 2005. DNA Vaccine, p. 257–289. *In* Non-Viral Vectors for Gene Therapy, Second Edition: Part 2. Elsevier.

23. Ellis RW. 1994. New Vaccine Technologies. JAMA 271:929.

24. Liu MA. 2003. DNA vaccines: a review. J Intern Med 253:402–410.

25. Kutzler MA, Weiner DB. 2008. DNA vaccines: ready for prime time? Nat Rev Genet 9:776–788.

26. Sternberg A, Naujokat C. 2020. Structural features of coronavirus SARS-CoV-2 spike protein: Targets for vaccination. Life Sciences 257:118056.

27. Li F. 2016. Structure, Function, and Evolution of Coronavirus Spike Proteins. Annual Review of Virology 3:237–261.

28. Kirchdoerfer RN, Cottrell CA, Wang N, Pallesen J, Yassine HM, Turner HL, Corbett KS, Graham BS, McLellan JS, Ward AB. 2016. Pre-fusion structure of a human coronavirus spike protein. Nature 531:118–121.

29. Walls AC, Park Y-J, Tortorici MA, Wall A, McGuire AT, Veesler D. 2020. Structure, Function, and Antigenicity of the SARS-CoV-2 Spike Glycoprotein. Cell 181:281–292.e6.

30. Liu C, Mendonça L, Yang Y, Gao Y, Shen C, Liu J, Ni T, Ju B, Liu C, Tang X, Wei J, Ma X, Zhu Y, Liu W, Xu S, Liu Y, Yuan J, Wu J, Liu Z, Zhang Z, Liu L, Wang P, Zhang P. 2020. The Architecture of Inactivated SARS-CoV-2 with Postfusion Spikes Revealed by Cryo-EM and Cryo-ET. Structure 28:1218–1224.e4.

31. Ke Z, Oton J, Qu K, Cortese M, Zila V, McKeane L, Nakane T, Zivanov J, Neufeldt CJ, Cerikan B, Lu JM, Peukes J, Xiong X, Kräusslich H-G, Scheres SHW, Bartenschlager R, Briggs JAG. 2020. Structures and distributions of SARS-CoV-2 spike proteins on intact virions. Nature 588:498–502.

32. Pallesen J, Wang N, Corbett KS, Wrapp D, Kirchdoerfer RN, Turner HL, Cottrell CA, Becker MM, Wang L, Shi W, Kong W-P, Andres EL, Kettenbach AN, Denison MR, Chappell JD, Graham BS, Ward AB, McLellan JS. 2017. Immunogenicity and structures of a rationally designed prefusion MERS-CoV spike antigen. Proc Natl Acad Sci USA 114.

33.     Belouzard S, Millet JK, Licitra BN, Whittaker GR. 2012. Mechanisms of Coronavirus Cell Entry Mediated by the Viral Spike Protein. Viruses 4:1011–1033.

34.     Jaimes JA, André NM, Chappie JS, Millet JK, Whittaker GR. 2020. Phylogenetic Analysis and Structural Modeling of SARS-CoV-2 Spike Protein Reveals an Evolutionary Distinct and Proteolytically Sensitive Activation Loop. Journal of Molecular Biology 432:3309–3325.

35.     Hsieh C-L, Goldsmith JA, Schaub JM, DiVenere AM, Kuo H-C, Javanmardi K, Le KC, Wrapp D, Lee AG, Liu Y, Chou C-W, Byrne PO, Hjorth CK, Johnson NV, Ludes-Meyers J, Nguyen AW, Park J, Wang N, Amengor D, Lavinder JJ, Ippolito GC, Maynard JA, Finkelstein IJ, McLellan JS. 2020. Structure-based design of prefusion-stabilized SARS-CoV-2 spikes. Science 369:1501–1505.

36.     Bos R, Rutten L, van der Lubbe JEM, Bakkers MJG, Hardenberg G, Wegmann F, Zuijdgeest D, de Wilde AH, Koornneef A, Verwilligen A, van Manen D, Kwaks T, Vogels R, Dalebout TJ, Myeni SK, Kikkert M, Snijder EJ, Li Z, Barouch DH, Vellinga J, Langedijk JPM, Zahn RC, Custers J, Schuitemaker H. 2020. Ad26 vector-based COVID-19 vaccine encoding a prefusion-stabilized SARS-CoV-2 Spike immunogen induces potent humoral and cellular immune responses. npj Vaccines 5.

37.     Marrack P, McKee AS, Munks MW. 2009. Towards an understanding of the adjuvant action of aluminium. Nature Reviews Immunology 9:287–293.

38.     Hayashi T, Momota M, Kuroda E, Kusakabe T, Kobari S, Makisaka K, Ohno Y, Suzuki Y, Nakagawa F, Lee MSJ, Coban C, Onodera R, Higashi T, Motoyama K, Ishii KJ, Arima H. 2018. DAMP-Inducing Adjuvant and PAMP Adjuvants Parallelly Enhance Protective Type-2 and Type-1 Immune Responses to Influenza Split Vaccination. Frontiers in Immunology 9:2619.

39.     Wang Z-B, Xu J. 2020. Better Adjuvants for Better Vaccines: Progress in Adjuvant Delivery Systems, Modifications, and Adjuvant–Antigen Codelivery. Vaccines 8:128.

40.     Rando HM, Greene CS, Robson MP, Boca SM, Wellhausen N, Lordan R, Brueffer C, Ray S, McGowan LD, Gitter A, Dattoli AA, Velazquez R, Barton JP, Field JM, Ramsundar B, MacLean AL, Lee AJ, Immunology Institute of the Icahn School of Medicine, Hu F, Jadavji NM, Sell E, Wang J, Rafizadeh DN, Skelly AN, Guebila MB, Kolla L, Manheim D, Ghosh S, Byrd JB, Park Y, Bansal V, Capone S, Dziak JJ, Sun Y, Qi Y, Shinholster L, Lukan T, Knyazev S, Perrin D, Mangul S, Das S, Szeto GL, Lubiana T, Mai D, COVID-19 Review Consortium, Goel RR, Boerckel JD, Naik A, Sun Y, Himmelstein DS, Kamil JP, Meyer JG, Mundo AI. 2023. SARS-CoV-2 and COVID-19: An Evolving Review of Diagnostics and Therapeutics.

41.     Hobernik D, Bros M. 2018. DNA Vaccines—How Far From Clinical Use? IJMS 19:3605.

42.     Ghaffarifar F. 2018. Plasmid DNA vaccines: where are we now? Drugs Today 54:315.

43.     Glenting J, Wessels S. 2005. Ensuring safety of DNA vaccines. Microb Cell Fact 4.

44. Williams J. 2013. Vector Design for Improved DNA Vaccine Efficacy, Safety and Production. Vaccines 1:225–249.

45. Lim M, Badruddoza AZM, Firdous J, Azad M, Mannan A, Al-Hilal TA, Cho C-S, Islam MA. 2020. Engineered Nanodelivery Systems to Improve DNA Vaccine Technologies. Pharmaceutics 12:30.

46. Types of Vaccines – COVID19 Vaccine Tracker. https://covid19.trackvaccines.org/types-of-vaccines/. Retrieved 5 December 2022.

47. DNA vaccines. World Health Organization. https://www.who.int/biologicals/areas/vaccines/dna/en. Retrieved 5 August 2022.

48. Lapuente D, Stab V, Storcksdieck genannt Bonsmann M, Maaske A, Köster M, Xiao H, Ehrhardt C, Tenbusch M. 2020. Innate signalling molecules as genetic adjuvants do not alter the efficacy of a DNA-based influenza A vaccine. PLoS ONE 15:e0231138.

49. Liu MA, Ulmer JB. 2005. Human Clinical Trials of Plasmid DNA Vaccines, p. 25–40. *In* Advances in Genetics. Elsevier.

50. Roy MJ, Wu MS, Barr LJ, Fuller JT, Tussey LG, Speller S, Culp J, Burkholder JK, Swain WF, Dixon RM, Widera G, Vessey R, King A, Ogg G, Gallimore A, Haynes JR, Heydenburg Fuller D. 2000. Induction of antigen-specific CD8+ T cells, T helper cells, and protective levels of antibody in humans by particle-mediated administration of a hepatitis B virus DNA vaccine. Vaccine 19:764–778.

51. Weiner DB, Nabel GJ. 2018. Development of Gene-Based Vectors for Immunization, p. 1305–1319.e8. *In* Plotkin's Vaccines. Elsevier.

52. A. Gómez L, A. Oñate A. 2019. Plasmid-Based DNA VaccinesPlasmid. IntechOpen.

53. Eusébio D, Neves AR, Costa D, Biswas S, Alves G, Cui Z, Sousa Â. 2021. Methods to improve the immunogenicity of plasmid DNA vaccines. Drug Discovery Today 26:2575–2592.

54. Garver K, LaPatra S, Kurath G. 2005. Efficacy of an infectious hematopoietic necrosis (IHN) virus DNA vaccine in Chinook Oncorhynchus tshawytscha and sockeye O. nerka salmon. Dis Aquat Org 64:13–22.

55. Thacker EL, Holtkamp DJ, Khan AS, Brown PA, Draghia-Akli R. 2006. Plasmid-mediated growth hormone-releasing hormone efficacy in reducing disease associated with Mycoplasma hyopneumoniae and porcine reproductive and respiratory syndrome virus infection1. Journal of Animal Science 84:733–742.

56. Davidson AH, Traub-Dargatz JL, Rodeheaver RM, Ostlund EN, Pedersen DD, Moorhead RG, Stricklin JB, Dewell RD, Roach SD, Long RE, Albers SJ, Callan RJ, Salman MD. 2005. Immunologic responses to West Nile virus in vaccinated and clinically affected horses. javma 226:240–245.

57.     Langellotti CA, Gammella M, Soria I, Bellusci C, Quattrocchi V, Vermeulen M, Mongini C, Zamorano PI. 2021. An Improved DNA Vaccine Against Bovine Herpesvirus-1 Using CD40L and a Chemical Adjuvant Induces Specific Cytotoxicity in Mice. Viral Immunology 34:68–78.

58.     Collins C, Lorenzen N, Collet B. 2019. DNA vaccination for finfish aquaculture. Fish & Shellfish Immunology 85:106–125.

59.     Chakraborty C, Agoramoorthy G. 2020. India's cost-effective COVID-19 vaccine development initiatives. Vaccine 38:7883–7884.

60.     Mallapaty S. 2021. India's DNA COVID vaccine is a world first – more are coming. Nature 597:161–162.

61.     Abbasi J. 2021. India's New COVID-19 DNA Vaccine for Adolescents and Adults Is a First. JAMA 326:1365.

62.     Safety, Tolerability and Immunogenicity of INO-4800 for COVID-19 in Healthy Volunteers - Full Text View - ClinicalTrials.gov. https://clinicaltrials.gov/ct2/show/NCT04336410. Retrieved 8 February 2021.

63.     Diehl MC, Lee JC, Daniels SE, Tebas P, Khan AS, Giffear M, Sardesai NY, Bagarazzi ML. 2013. Tolerability of intramuscular and intradermal delivery by CELLECTRA® adaptive constant current electroporation device in healthy volunteers. Human Vaccines & Immunotherapeutics 9:2246–2252.

64.     Sardesai NY, Weiner DB. 2011. Electroporation delivery of DNA vaccines: prospects for success. Current Opinion in Immunology 23:421–429.

65.     Tebas P, Yang S, Boyer JD, Reuschel EL, Patel A, Christensen-Quick A, Andrade VM, Morrow MP, Kraynyak K, Agnes J, Purwar M, Sylvester A, Pawlicki J, Gillespie E, Maricic I, Zaidi FI, Kim KY, Dia Y, Frase D, Pezzoli P, Schultheis K, Smith TRF, Ramos SJ, McMullan T, Buttigieg K, Carroll MW, Ervin J, Diehl MC, Blackwood E, Mammen MP, Lee J, Dallas MJ, Brown AS, Shea JE, Kim JJ, Weiner DB, Broderick KE, Humeau LM. 2021. Safety and immunogenicity of INO-4800 DNA vaccine against SARS-CoV-2: A preliminary report of an open-label, Phase 1 clinical trial. EClinicalMedicine 31:100689.

66.     Mammen MP Jr., Tebas P, Agnes J, Giffear M, Kraynyak KA, Blackwood E, Amante D, Reuschel EL, Purwar M, Christensen-Quick A, Liu N, Andrade VM, Carter J, Garufi G, Diehl MC, Sylvester A, Morrow MP, Pezzoli P, Kulkarni AJ, Zaidi FI, Frase D, Liaw K, Badie H, Simon KO, Smith TRF, Ramos S, Spitz R, Juba RJ, Lee J, Dallas M, Brown AS, Shea JE, Kim JJ, Weiner DB, Broderick KE, Boyer JD, Humeau LM. 2021. Safety and immunogenicity of INO-4800 DNA vaccine against SARS-CoV-2: a preliminary report of a randomized, blinded, placebo-controlled, Phase 2 clinical trial in adults at high risk of viral exposure. Cold Spring Harbor Laboratory.

67.     INOVIO Receives U.S. FDA Authorization to Proceed with INNOVATE Phase 3 Segment for its COVID-19 Vaccine Candidate, INO-4800, in the U.S. https://ir.inovio.com/news-releases/news-releases-details/2021/INOVIO-Receives-U.S.-FDA-Authorization-to-Proceed-

with-INNOVATE-Phase-3-Segment-for-its-COVID-19-Vaccine-Candidate-INO-4800-in-the-U.S/default.aspx. Retrieved 5 December 2022.

68. INOVIO Further Expands INNOVATE Phase 3 Trial for COVID-19 DNA Vaccine Candidate INO-4800 With Regulatory Authorization from India. https://ir.inovio.com/news-releases/news-releases-details/2021/INOVIO-Further-Expands-INNOVATE-Phase-3-Trial-for-COVID-19-DNA-Vaccine-Candidate-INO-4800-With-Regulatory-Authorization-from-India/default.aspx. Retrieved 5 December 2022.

69. INOVIO Expands INNOVATE Phase 3 for INO-4800, its DNA Vaccine Candidate for COVID-19, to include Colombia following Regulatory Authorization. https://ir.inovio.com/news-releases/news-releases-details/2021/INOVIO-Expands-INNOVATE-Phase-3-for-INO-4800-its-DNA-Vaccine-Candidate-for-COVID-19-to-include-Colombia-following-Regulatory-Authorization/default.aspx. Retrieved 5 December 2022.

70. INOVIO Receives Regulatory Authorization to Conduct Phase 3 Efficacy Trial of its COVID-19 DNA Vaccine Candidate, INO-4800, in Mexico. https://ir.inovio.com/news-releases/news-releases-details/2021/INOVIO-Receives-Regulatory-Authorization-to-Conduct-Phase-3-Efficacy-Trial-of-its-COVID-19-DNA-Vaccine-Candidate-INO-4800-in-Mexico/default.aspx. Retrieved 5 December 2022.

71. Momin T, Kansagra K, Patel H, Sharma S, Sharma B, Patel J, Mittal R, Sanmukhani J, Maithal K, Dey A, Chandra H, Rajanathan CT, Pericherla HP, Kumar P, Narkhede A, Parmar D. 2021. Safety and Immunogenicity of a DNA SARS-CoV-2 vaccine (ZyCoV-D): Results of an open-label, non-randomized phase I part of phase I/II clinical study by intradermal route in healthy subjects in India. EClinicalMedicine 38:101020.

72. Khobragade A, Bhate S, Ramaiah V, Deshpande S, Giri K, Phophle H, Supe P, Godara I, Revanna R, Nagarkar R, Sanmukhani J, Dey A, Rajanathan TMC, Kansagra K, Koradia P. 2022. Efficacy, safety, and immunogenicity of the DNA SARS-CoV-2 vaccine (ZyCoV-D): the interim efficacy results of a phase 3, randomised, double-blind, placebo-controlled study in India. The Lancet 399:1313–1321.

73. Zydus Cadila: ZyCoV-D – COVID19 Vaccine Tracker. https://covid19.trackvaccines.org/vaccines/29/. Retrieved 5 December 2022.

74. Mathieu E, Ritchie H, Ortiz-Ospina E, Roser M, Hasell J, Appel C, Giattino C, Rodés-Guirao L. 2021. A global database of COVID-19 vaccinations. Nat Hum Behav 5:947–953.

75. 2021. Covishield, Covaxin, ZyCoV-D Makers to Assess Efficacy of Vaccine as They Await Data on Omicron. News18. https://www.news18.com/news/india/covishield-covaxin-zycov-d-makers-to-assess-efficacy-of-their-vaccine-as-they-await-data-on-omicron-4504295.html. Retrieved 5 December 2022.

76. Andrade VM, Christensen-Quick A, Agnes J, Tur J, Reed C, Kalia R, Marrero I, Elwood D, Schultheis K, Purwar M, Reuschel E, McMullan T, Pezzoli P, Kraynyak K, Sylvester A, Mammen MP, Tebas P, Joseph Kim J, Weiner DB, Smith TRF, Ramos SJ, Humeau LM, Boyer

JD, Broderick KE. 2021. INO-4800 DNA vaccine induces neutralizing antibodies and T cell activity against global SARS-CoV-2 variants. npj Vaccines 6.

77. Kraynyak KA, Blackwood E, Agnes J, Tebas P, Giffear M, Amante D, Reuschel EL, Purwar M, Christensen-Quick A, Liu N, Andrade VM, Diehl MC, Wani S, Lupicka M, Sylvester A, Morrow MP, Pezzoli P, McMullan T, Kulkarni AJ, Zaidi FI, Frase D, Liaw K, Smith TRF, Ramos SJ, Ervin J, Adams M, Lee J, Dallas M, Shah Brown A, Shea JE, Kim JJ, Weiner DB, Broderick KE, Humeau LM, Boyer JD, Mammen MP Jr. 2022. SARS-CoV-2 DNA Vaccine INO-4800 Induces Durable Immune Responses Capable of Being Boosted in a Phase 1 Open-Label Trial. The Journal of Infectious Diseases 225:1923–1932.

78. Inc IP. INOVIO Announces Strategy to Address Omicron (B.1.1.529) and Future SARS-CoV-2 Variants. https://www.prnewswire.com/news-releases/inovio-announces-strategy-to-address-omicron-b1-1-529-and-future-sars-cov-2-variants-301433776.html. Retrieved 5 December 2022.

79. Walters JN, Schouest B, Patel A, Reuschel EL, Schultheis K, Parzych E, Maricic I, Gary EN, Purwar M, Andrade VM, Doan A, Elwood D, Eblimit Z, Nguyen B, Frase D, Zaidi FI, Kulkarni A, Generotti A, Joseph Kim J, Humeau LM, Ramos SJ, Smith TRF, Weiner DB, Broderick KE. 2022. Prime-boost vaccination regimens with INO-4800 and INO-4802 augment and broaden immune responses against SARS-CoV-2 in nonhuman primates. Vaccine 40:2960–2969.

80. Reed CC, Schultheis K, Andrade VM, Kalia R, Tur J, Schouest B, Elwood D, Walters JN, Maricic I, Doan A, Vazquez M, Eblimit Z, Pezzoli P, Amante D, Porto M, Narvaez B, Lok M, Spence B, Bradette H, Horn H, Yang M, Fader J, Ferrer R, Weiner DB, Kar S, Kim JJ, Humeau LM, Ramos SJ, Smith TRF, Broderick KE. 2021. Design, immunogenicity and efficacy of a Pan-SARS-CoV-2 synthetic DNA vaccine. Cold Spring Harbor Laboratory.

81. Lauer KB, Borrow R, Blanchard TJ. 2017. Multivalent and Multipathogen Viral Vector Vaccines. Clin Vaccine Immunol 24.

82. Ewer KJ, Lambe T, Rollier CS, Spencer AJ, Hill AV, Dorrell L. 2016. Viral vectors as vaccine platforms: from immunogenicity to impact. Current Opinion in Immunology 41:47–54.

83. Antrobus RD, Coughlan L, Berthoud TK, Dicks MD, Hill AV, Lambe T, Gilbert SC. 2014. Clinical Assessment of a Novel Recombinant Simian Adenovirus ChAdOx1 as a Vectored Vaccine Expressing Conserved Influenza A Antigens. Molecular Therapy 22:668–674.

84. Al-Kassmy J, Pedersen J, Kobinger G. 2020. Vaccine Candidates against Coronavirus Infections. Where Does COVID-19 Stand? Viruses 12:861.

85. Pastoret P-P, Vanderplasschen A. 2003. Poxviruses as vaccine vectors. Comparative Immunology, Microbiology and Infectious Diseases 26:343–355.

86. García-Arriaza J, Esteban M. 2014. Enhancing poxvirus vectors vaccine immunogenicity. Human Vaccines & Immunotherapeutics 10:2235–2244.

87.     Lasaro MO, Ertl HC. 2009. New Insights on Adenovirus as Vaccine Vectors. Molecular Therapy 17:1333–1339.

88.     Roberts A, Buonocore L, Price R, Forman J, Rose JK. 1999. Attenuated vesicular stomatitis viruses as vaccine vectors. J Virol 73:3723–32.

89.     Lichty BD, Power AT, Stojdl DF, Bell JC. 2004. Vesicular stomatitis virus: re-inventing the bullet. Trends in Molecular Medicine 10:210–216.

90.     Rollier CS, Reyes-Sandoval A, Cottingham MG, Ewer K, Hill AV. 2011. Viral vectors as vaccine platforms: deployment in sight. Current Opinion in Immunology 23:377–382.

91.     Nayak S, Herzog RW. 2009. Progress and prospects: immune responses to viral vectors. Gene Ther 17:295–304.

92.     Ura T, Okuda K, Shimada M. 2014. Developments in Viral Vector-Based Vaccines. Vaccines 2:624–641.

93.     Moffatt S, Hays J, HogenEsch H, Mittal SK. 2000. Circumvention of Vector-Specific Neutralizing Antibody Response by Alternating Use of Human and Non-Human Adenoviruses: Implications in Gene Therapy. Virology 272:159–167.

94.     Fausther-Bovendo H, Kobinger GP. 2014. Pre-existing immunity against Ad vectors. Human Vaccines & Immunotherapeutics 10:2875–2884.

95.     Vrba SM, Kirk NM, Brisse ME, Liang Y, Ly H. 2020. Development and Applications of Viral Vectored Vaccines to Combat Zoonotic and Emerging Public Health Threats. Vaccines 8:680.

96.     Ollmann Saphire E. 2020. A Vaccine against Ebola Virus. Cell 181:6.

97.     Bliss CM, Drammeh A, Bowyer G, Sanou GS, Jagne YJ, Ouedraogo O, Edwards NJ, Tarama C, Ouedraogo N, Ouedraogo M, Njie-Jobe J, Diarra A, Afolabi MO, Tiono AB, Yaro JB, Adetifa UJ, Hodgson SH, Anagnostou NA, Roberts R, Duncan CJA, Cortese R, Viebig NK, Leroy O, Lawrie AM, Flanagan KL, Kampmann B, Imoukhuede EB, Sirima SB, Bojang K, Hill AVS, Nébié I, Ewer KJ. 2017. Viral Vector Malaria Vaccines Induce High-Level T Cell and Antibody Responses in West African Children and Infants. Molecular Therapy 25:547–559.

98.     Li S, Locke E, Bruder J, Clarke D, Doolan DL, Havenga MJE, Hill AVS, Liljestrom P, Monath TP, Naim HY, Ockenhouse C, Tang DC, Van Kampen KR, Viret J-F, Zavala F, Dubovsky F. 2007. Viral vectors for malaria vaccine development. Vaccine 25:2567–2574.

99.     Ledgerwood JE, DeZure AD, Stanley DA, Coates EE, Novik L, Enama ME, Berkowitz NM, Hu Z, Joshi G, Ploquin A, Sitar S, Gordon IJ, Plummer SA, Holman LA, Hendel CS, Yamshchikov G, Roman F, Nicosia A, Colloca S, Cortese R, Bailer RT, Schwartz RM, Roederer M, Mascola JR, Koup RA, Sullivan NJ, Graham BS. 2017. Chimpanzee Adenovirus Vector Ebola Vaccine. N Engl J Med 376:928–938.

100.    Geisbert TW, Feldmann H. 2011. Recombinant Vesicular Stomatitis Virus–Based Vaccines Against Ebola and Marburg Virus Infections. The Journal of Infectious Diseases 204:S1075–S1081.

101.    Marzi A, Feldmann H. 2014. Ebola virus vaccines: an overview of current approaches. Expert Review of Vaccines 13:521–531.

102.    Parks CL, Picker LJ, King CR. 2013. Development of replication-competent viral vectors for HIV vaccine delivery. Current Opinion in HIV and AIDS 8:402–411.

103.    Trivedi S, Jackson RJ, Ranasinghe C. 2014. Different HIV pox viral vector-based vaccines and adjuvants can induce unique antigen presenting cells that modulate CD8 T cell avidity. Virology 468-470:479–489.

104.    See RH, Zakhartchouk AN, Petric M, Lawrence DJ, Mok CPY, Hogan RJ, Rowe T, Zitzow LA, Karunakaran KP, Hitt MM, Graham FL, Prevec L, Mahony JB, Sharon C, Auperin TC, Rini JM, Tingle AJ, Scheifele DW, Skowronski DM, Patrick DM, Voss TG, Babiuk LA, Gauldie J, Roper RL, Brunham RC, Finlay BB. 2006. Comparative evaluation of two severe acute respiratory syndrome (SARS) vaccine candidates in mice challenged with SARS coronavirus. Journal of General Virology 87:641–650.

105.    See RH, Petric M, Lawrence DJ, Mok CPY, Rowe T, Zitzow LA, Karunakaran KP, Voss TG, Brunham RC, Gauldie J, Finlay BB, Roper RL. 2008. Severe acute respiratory syndrome vaccine efficacy in ferrets: whole killed virus and adenovirus-vectored vaccines. Journal of General Virology 89:2136–2146.

106.    Yeung HT. 2018. Update with the development of Ebola vaccines and implications of emerging evidence to inform future policy recommendations. World Health Organization SAGE meeting background. https://www.who.int/immunization/sage/meetings/2018/october/2_Ebola_SAGE2018Oct_BgDoc_20180919.pdf.

107.    Alharbi NK, Padron-Regalado E, Thompson CP, Kupke A, Wells D, Sloan MA, Grehan K, Temperton N, Lambe T, Warimwe G, Becker S, Hill AVS, Gilbert SC. 2017. ChAdOx1 and MVA based vaccine candidates against MERS-CoV elicit neutralising antibodies and cellular immune responses in mice. Vaccine 35:3780–3788.

108.    van Doremalen N, Haddock E, Feldmann F, Meade-White K, Bushmaker T, Fischer RJ, Okumura A, Hanley PW, Saturday G, Edwards NJ, Clark MHA, Lambe T, Gilbert SC, Munster VJ. 2020. A single dose of ChAdOx1 MERS provides protective immunity in rhesus macaques. Sci Adv 6.

109.    Folegatti PM, Bittaye M, Flaxman A, Lopez FR, Bellamy D, Kupke A, Mair C, Makinson R, Sheridan J, Rohde C, Halwe S, Jeong Y, Park Y-S, Kim J-O, Song M, Boyd A, Tran N, Silman D, Poulton I, Datoo M, Marshall J, Themistocleous Y, Lawrie A, Roberts R, Berrie E, Becker S, Lambe T, Hill A, Ewer K, Gilbert S. 2020. Safety and immunogenicity of a candidate [Middle East respiratory syndrome coronavirus viral-vectored vaccine: a dose-escalation, open-label, non-randomised, uncontrolled, phase 1 trial](). The Lancet Infectious Diseases 20:816–826.

110. Jordahl K, Van Den Bossche J, Fleischmann M, McBride J, Wasserman J, Badaracco AG, Gerard J, Snow AD, Tratner J, Perry M, Farmer C, Hjelle GA, Cochran M, Gillies S, Culbertson L, Bartos M, Ward B, Caria G, Taves M, Eubank N, Sangarshanan, Flavin J, Richards M, Rey S, Maxalbert, Bilogur A, Ren C, Arribas-Bel D, Mesejo-León D, Wasser L. 2021. geopandas/geopandas: v0.10.2 (v0.10.2). Zenodo. https://doi.org/gqkzpv.

111. van Doremalen N, Lambe T, Spencer A, Belij-Rammerstorfer S, Purushotham JN, Port JR, Avanzato VA, Bushmaker T, Flaxman A, Ulaszewska M, Feldmann F, Allen ER, Sharpe H, Schulz J, Holbrook M, Okumura A, Meade-White K, Pérez-Pérez L, Edwards NJ, Wright D, Bissett C, Gilbride C, Williamson BN, Rosenke R, Long D, Ishwarbhai A, Kailath R, Rose L, Morris S, Powers C, Lovaglio J, Hanley PW, Scott D, Saturday G, de Wit E, Gilbert SC, Munster VJ. 2020. [ChAdOx1 nCoV-19 vaccine prevents SARS-CoV-2 pneumonia in rhesus macaques](). Nature 586:578–582.

112. Folegatti PM, Ewer KJ, Aley PK, Angus B, Becker S, Belij-Rammerstorfer S, Bellamy D, Bibi S, Bittaye M, Clutterbuck EA, Dold C, Faust SN, Finn A, Flaxman AL, Hallis B, Heath P, Jenkin D, Lazarus R, Makinson R, Minassian AM, Pollock KM, Ramasamy M, Robinson H, Snape M, Tarrant R, Voysey M, Green C, Douglas AD, Hill AVS, Lambe T, Gilbert SC, Pollard AJ, Aboagye J, Adams K, Ali A, Allen E, Allison JL, Anslow R, Arbe-Barnes EH, Babbage G, Baillie K, Baker M, Baker N, Baker P, Baleanu I, Ballaminut J, Barnes E, Barrett J, Bates L, Batten A, Beadon K, Beckley R, Berrie E, Berry L, Beveridge A, Bewley KR, Bijker EM, Bingham T, Blackwell L, Blundell CL, Bolam E, Boland E, Borthwick N, Bower T, Boyd A, Brenner T, Bright PD, Brown-O'Sullivan C, Brunt E, Burbage J, Burge S, Buttigieg KR, Byard N, Cabera Puig I, Calvert A, Camara S, Cao M, Cappuccini F, Carr M, Carroll MW, Carter V, Cathie K, Challis RJ, Charlton S, Chelysheva I, Cho J-S, Cicconi P, Cifuentes L, Clark H, Clark E, Cole T, Colin-Jones R, Conlon CP, Cook A, Coombes NS, Cooper R, Cosgrove CA, Coy K, Crocker WEM, Cunningham CJ, Damratoski BE, Dando L, Datoo MS, Davies H, De Graaf H, Demissie T, Di Maso C, Dietrich I, Dong T, Donnellan FR, Douglas N, Downing C, Drake J, Drake-Brockman R, Drury RE, Dunachie SJ, Edwards NJ, Edwards FDL, Edwards CJ, Elias SC, Elmore MJ, Emary KRW, English MR, Fagerbrink S, Felle S, Feng S, Field S, Fixmer C, Fletcher C, Ford KJ, Fowler J, Fox P, Francis E, Frater J, Furze J, Fuskova M, Galiza E, Gbesemete D, Gilbride C, Godwin K, Gorini G, Goulston L, Grabau C, Gracie L, Gray Z, Guthrie LB, Hackett M, Halwe S, Hamilton E, Hamlyn J, Hanumunthadu B, Harding I, Harris SA, Harris A, Harrison D, Harrison C, Hart TC, Haskell L, Hawkins S, Head I, Henry JA, Hill J, Hodgson SHC, Hou MM, Howe E, Howell N, Hutlin C, Ikram S, Isitt C, Iveson P, Jackson S, Jackson F, James SW, Jenkins M, Jones E, Jones K, Jones CE, Jones B, Kailath R, Karampatsas K, Keen J, Kelly S, Kelly D, Kerr D, Kerridge S, Khan L, Khan U, Killen A, Kinch J, King TB, King L, King J, Kingham-Page L, Klenerman P, Knapper F, Knight JC, Knott D, Koleva S, Kupke A, Larkworthy CW, Larwood JPJ, Laskey A, Lawrie AM, Lee A, Ngan Lee KY, Lees EA, Legge H, Lelliott A, Lemm N-M, Lias AM, Linder A, Lipworth S, Liu X, Liu S, Lopez Ramon R, Lwin M, Mabesa F, Madhavan M, Mallett G, Mansatta K, Marcal I, Marinou S, Marlow E, Marshall JL, Martin J, McEwan J, McInroy L, Meddaugh G, Mentzer AJ, Mirtorabi N, Moore M, Moran E, Morey E, Morgan V, Morris SJ, Morrison H, Morshead G, Morter R, Mujadidi YF, Muller J, Munera-Huertas T, Munro C, Munro A, Murphy S, Munster VJ, Mweu P, Noé A, Nugent FL,

Nuthall E, O'Brien K, O'Connor D, Oguti B, Oliver JL, Oliveira C, O'Reilly PJ, Osborn M, Osborne P, Owen C, Owens D, Owino N, Pacurar M, Parker K, Parracho H, Patrick-Smith M, Payne V, Pearce J, Peng Y, Peralta Alvarez MP, Perring J, Pfafferott K, Pipini D, Plested E, Pluess-Hall H, Pollock K, Poulton I, Presland L, Provstgaard-Morys S, Pulido D, Radia K, Ramos Lopez F, Rand J, Ratcliffe H, Rawlinson T, Rhead S, Riddell A, Ritchie AJ, Roberts H, Robson J, Roche S, Rohde C, Rollier CS, Romani R, Rudiansyah I, Saich S, Sajjad S, Salvador S, Sanchez Riera L, Sanders H, Sanders K, Sapaun S, Sayce C, Schofield E, Screaton G, Selby B, Semple C, Sharpe HR, Shaik I, Shea A, Shelton H, Silk S, Silva-Reyes L, Skelly DT, Smee H, Smith CC, Smith DJ, Song R, Spencer AJ, Stafford E, Steele A, Stefanova E, Stockdale L, Szigeti A, Tahiri-Alaoui A, Tait M, Talbot H, Tanner R, Taylor IJ, Taylor V, Te Water Naude R, Thakur N, Themistocleous Y, Themistocleous A, Thomas M, Thomas TM, Thompson A, Thomson-Hill S, Tomlins J, Tonks S, Towner J, Tran N, Tree JA, Truby A, Turkentine K, Turner C, Turner N, Turner S, Tuthill T, Ulaszewska M, Varughese R, Van Doremalen N, Veighey K, Verheul MK, Vichos I, Vitale E, Walker L, Watson MEE, Welham B, Wheat J, White C, White R, Worth AT, Wright D, Wright S, Yao XL, Yau Y. 2020. Safety and immunogenicity of the ChAdOx1 nCoV-19 vaccine against SARS-CoV-2: a preliminary report of a phase 1/2, single-blind, randomised controlled trial. The Lancet 396:467–478.

113. Jones I, Roy P. 2021. Sputnik V COVID-19 vaccine candidate appears safe and effective. The Lancet 397:642–643.

114. Barouch DH, Kik SV, Weverling GJ, Dilan R, King SL, Maxfield LF, Clark S, Ng'ang'a D, Brandariz KL, Abbink P, Sinangil F, de Bruyn G, Gray GE, Roux S, Bekker L-G, Dilraj A, Kibuuka H, Robb ML, Michael NL, Anzala O, Amornkul PN, Gilmour J, Hural J, Buchbinder SP, Seaman MS, Dolin R, Baden LR, Carville A, Mansfield KG, Pau MG, Goudsmit J. 2011. International seroepidemiology of adenovirus serotypes 5, 26, 35, and 48 in pediatric and adult populations. Vaccine 29:5203–5209.

115. Office USGA. Operation Warp Speed: Accelerated COVID-19 Vaccine Development Status and Efforts to Address Manufacturing Challenges. https://www.gao.gov/products/gao-21-319. Retrieved 5 December 2022.

116. Johnson & Johnson Announces a Lead Vaccine Candidate for COVID-19; Landmark New Partnership with U.S. Department of Health & Human Services; and Commitment to Supply One Billion Vaccines Worldwide for Emergency Pandemic Use | Johnson & Johnson. Content Lab US. https://www.jnj.com/johnson-johnson-announces-a-lead-vaccine-candidate-for-covid-19-landmark-new-partnership-with-u-s-department-of-health-human-services-and-commitment-to-supply-one-billion-vaccines-worldwide-for-emergency-pandemic-use. Retrieved 5 December 2022.

117. Sadoff J, Le Gars M, Shukarev G, Heerwegh D, Truyers C, de Groot AM, Stoop J, Tete S, Van Damme W, Leroux-Roels I, Berghmans P-J, Kimmel M, Van Damme P, de Hoon J, Smith W, Stephenson KE, De Rosa SC, Cohen KW, McElrath MJ, Cormier E, Scheper G, Barouch DH, Hendriks J, Struyf F, Douoguih M, Van Hoof J, Schuitemaker H. 2021. Interim Results of a Phase 1–2a Trial of Ad26.COV2.S Covid-19 Vaccine. N Engl J Med 384:1824–1835.

118. Mercado NB, Zahn R, Wegmann F, Loos C, Chandrashekar A, Yu J, Liu J, Peter L, McMahan K, Tostanoski LH, He X, Martinez DR, Rutten L, Bos R, van Manen D, Vellinga J, Custers J, Langedijk JP, Kwaks T, Bakkers MJG, Zuijdgeest D, Rosendahl Huber SK, Atyeo C, Fischinger S, Burke JS, Feldman J, Hauser BM, Caradonna TM, Bondzie EA, Dagotto G, Gebre MS, Hoffman E, Jacob-Dolan C, Kirilova M, Li Z, Lin Z, Mahrokhian SH, Maxfield LF, Nampanya F, Nityanandam R, Nkolola JP, Patel S, Ventura JD, Verrington K, Wan H, Pessaint L, Van Ry A, Blade K, Strasbaugh A, Cabus M, Brown R, Cook A, Zouantchangadou S, Teow E, Andersen H, Lewis MG, Cai Y, Chen B, Schmidt AG, Reeves RK, Baric RS, Lauffenburger DA, Alter G, Stoffels P, Mammen M, Van Hoof J, Schuitemaker H, Barouch DH. 2020. Single-shot Ad26 vaccine protects against SARS-CoV-2 in rhesus macaques. Nature 586:583–588.

119. Tostanoski LH, Wegmann F, Martinot AJ, Loos C, McMahan K, Mercado NB, Yu J, Chan CN, Bondoc S, Starke CE, Nekorchuk M, Busman-Sahay K, Piedra-Mora C, Wrijil LM, Ducat S, Custers J, Atyeo C, Fischinger S, Burke JS, Feldman J, Hauser BM, Caradonna TM, Bondzie EA, Dagotto G, Gebre MS, Jacob-Dolan C, Lin Z, Mahrokhian SH, Nampanya F, Nityanandam R, Pessaint L, Porto M, Ali V, Benetiene D, Tevi K, Andersen H, Lewis MG, Schmidt AG, Lauffenburger DA, Alter G, Estes JD, Schuitemaker H, Zahn R, Barouch DH. 2020. Ad26 vaccine protects against SARS-CoV-2 severe clinical disease in hamsters. Nat Med 26:1694–1700.

120. Solforosi L, Kuipers H, Huber SKR, van der Lubbe JEM, Dekking L, Czapska-Casey DN, Gil AI, Baert MRM, Drijver J, Vaneman J, van Huizen E, Choi Y, Vreugdenhil J, Kroos S, de Wilde AH, Kourkouta E, Custers J, Dalebout TJ, Myeni SK, Kikkert M, Snijder EJ, Barouch DH, Böszörményi KP, Stammes MA, Kondova I, Verschoor EJ, Verstrepen BE, Koopman G, Mooij P, Bogers WMJM, van Heerden M, Muchene L, Tolboom JTBM, Roozendaal R, Schuitemaker H, Wegmann F, Zahn RC. 2020. Immunogenicity and protective efficacy of one- and two-dose regimens of the Ad26.COV2.S COVID-19 vaccine candidate in adult and aged rhesus macaques. Cold Spring Harbor Laboratory.

121. Roozendaal R, Solforosi L, Stieh D, Serroyen J, Straetemans R, Wegmann F, Rosendahl Huber SK, van der Lubbe JEM, Hendriks J, le Gars M, Dekking L, Czapska-Casey DN, Guimera N, Janssen S, Tete S, Chandrashekar A, Mercado N, Yu J, Koudstaal W, Sadoff J, Barouch DH, Schuitemaker H, Zahn R. 2021. SARS-CoV-2 binding and neutralizing antibody levels after vaccination with Ad26.COV2.S predict durable protection in rhesus macaques. Cold Spring Harbor Laboratory.

122. Sadoff J, Gray G, Vandebosch A, Cárdenas V, Shukarev G, Grinsztejn B, Goepfert PA, Truyers C, Van Dromme I, Spiessens B, Vingerhoets J, Custers J, Scheper G, Robb ML, Treanor J, Ryser MF, Barouch DH, Swann E, Marovich MA, Neuzil KM, Corey L, Stoddard J, Hardt K, Ruiz-Guiñazú J, Le Gars M, Schuitemaker H, Van Hoof J, Struyf F, Douoguih M. 2022. Final Analysis of Efficacy and Safety of Single-Dose Ad26.COV2.S. N Engl J Med 386:847–860.

123. AstraZeneca's COVID-19 vaccine authorised for emergency supply in the UK. https://www.astrazeneca.com/media-centre/press-releases/2020/astrazenecas-covid-19-vaccine-authorised-in-uk.html. Retrieved 5 December 2022.

124.	The Brussels Times. https://www.brusselstimes.com/news-contents/world/149039/1-5-million-people-have-received-sputnik-v-vaccine-russia-says-russian-direct-investment-fund-mikhail-murashko. Retrieved 5 December 2022.

125.	Daventry M. 2021. Hungary becomes first EU country to deploy Russia's COVID-19 vaccine. euronews. https://www.euronews.com/2021/02/12/hungary-to-begin-using-russia-s-sputnik-v-vaccine-today. Retrieved 5 December 2022.

126.	2021. San Marino buys Russia's Sputnik V after EU vaccine delivery delays. euronews. https://www.euronews.com/2021/02/24/san-marino-buys-russia-s-sputnik-v-after-eu-vaccine-delivery-delays. Retrieved 5 December 2022.

127.	AFP. 2020. Belarus Starts Coronavirus Vaccination With Sputnik V. The Moscow Times. https://www.themoscowtimes.com/2020/12/29/belarus-starts-coronavirus-vaccination-with-sputnik-v-a72512. Retrieved 5 December 2022.

128.	Logunov DY, Dolzhikova IV, Shcheblyakov DV, Tukhvatulin AI, Zubkova OV, Dzharullaeva AS, Kovyrshina AV, Lubenets NL, Grousova DM, Erokhova AS, Botikov AG, Izhaeva FM, Popova O, Ozharovskaya TA, Esmagambetov IB, Favorskaya IA, Zrelkin DI, Voronina DV, Shcherbinin DN, Semikhin AS, Simakova YV, Tokarskaya EA, Egorova DA, Shmarov MM, Nikitenko NA, Gushchin VA, Smolyarchuk EA, Zyryanov SK, Borisevich SV, Naroditsky BS, Gintsburg AL. 2021. Safety and efficacy of an rAd26 and rAd5 vector-based heterologous prime-boost COVID-19 vaccine: an interim analysis of a randomised controlled phase 3 trial in Russia. The Lancet 397:671–681.

129.	Sadoff J, Gray G, Vandebosch A, Cárdenas V, Shukarev G, Grinsztejn B, Goepfert PA, Truyers C, Fennema H, Spiessens B, Offergeld K, Scheper G, Taylor KL, Robb ML, Treanor J, Barouch DH, Stoddard J, Ryser MF, Marovich MA, Neuzil KM, Corey L, Cauwenberghs N, Tanner T, Hardt K, Ruiz-Guiñazú J, Le Gars M, Schuitemaker H, Van Hoof J, Struyf F, Douoguih M. 2021. Safety and Efficacy of Single-Dose Ad26.COV2.S Vaccine against Covid-19. N Engl J Med 384:2187–2201.

130.	2021. Janssen Investigational COVID-19 Vaccine: Interim Analysis of Phase 3 Clinical Data Released. National Institutes of Health (NIH). https://www.nih.gov/news-events/news-releases/janssen-investigational-covid-19-vaccine-interim-analysis-phase-3-clinical-data-released. Retrieved 5 December 2022.

131.	2021. Johnson & Johnson Announces Single-Shot Janssen COVID-19 Vaccine Candidate Met Primary Endpoints in Interim Analysis of its Phase 3 ENSEMBLE Trial. Janssen. https://www.janssen.com/emea/sites/www_janssen_com_emea/files/johnson_johnson_announces_single-shot_janssen_covid-19_vaccine_candidate_met_primary_endpoints_in_interim_analysis_of_its_phase_3_ensemble_trial.pdf.

132.	Burki TK. 2020. The Russian vaccine for COVID-19. The Lancet Respiratory Medicine 8:e85–e86.

133. Cohen J. 2020. Russia's claim of a successful COVID-19 vaccine doesn't pass the 'smell test,' critics say. Science https://doi.org/10.1126/science.abf6791.

134. Callaway E. 2020. Russia announces positive COVID-vaccine results from controversial trial. Nature https://doi.org/10.1038/d41586-020-03209-0.

135. Thorp HH. 2020. A dangerous rush for vaccines. Science 369:885–885.

136. Jr BL. Scientists worry whether Russia's 'Sputnik V' coronavirus vaccine is safe and effective. CNBC. https://www.cnbc.com/2020/08/11/scientists-worry-whether-russias-sputnik-v-coronavirus-vaccine-is-safe-and-effective.html. Retrieved 5 December 2022.

137. Logunov DY, Dolzhikova IV, Zubkova OV, Tukhvatulin AI, Shcheblyakov DV, Dzharullaeva AS, Grousova DM, Erokhova AS, Kovyrshina AV, Botikov AG, Izhaeva FM, Popova O, Ozharovskaya TA, Esmagambetov IB, Favorskaya IA, Zrelkin DI, Voronina DV, Shcherbinin DN, Semikhin AS, Simakova YV, Tokarskaya EA, Lubenets NL, Egorova DA, Shmarov MM, Nikitenko NA, Morozova LF, Smolyarchuk EA, Kryukov EV, Babira VF, Borisevich SV, Naroditsky BS, Gintsburg AL. 2020. Safety and immunogenicity of an rAd26 and rAd5 vector-based heterologous prime-boost COVID-19 vaccine in two formulations: two open, non-randomised phase 1/2 studies from Russia. The Lancet 396:887–897.

138. Clinical Trials. https://sputnikvaccine.com/about-vaccine/clinical-trials/. Retrieved 5 December 2022.

139. Phillips N, Cyranoski D, Mallapaty S. 2020. A leading coronavirus vaccine trial is on hold: scientists react. Nature https://doi.org/10.1038/d41586-020-02594-w.

140. Cyranoski D, Mallapaty S. 2020. Scientists relieved as coronavirus vaccine trial restarts — but question lack of transparency. Nature 585:331–332.

141. Robbins R, LaFraniere S, Weiland N, Kirkpatrick DD, Mueller B. 2020. Blunders Eroded U.S. Confidence in Early Vaccine Front-Runner. The New York Times.

142. 2020. Oxford/AstraZeneca Covid vaccine 'dose error' explained. BBC News.

143. Sanchez S, Palacio N, Dangi T, Ciucci T, Penaloza-MacMaster P. 2021. Fractionating a COVID-19 Ad5-vectored vaccine improves virus-specific immunity. Sci Immunol 6.

144. Wolf ME, Luz B, Niehaus L, Bhogal P, Bäzner H, Henkes H. 2021. Thrombocytopenia and Intracranial Venous Sinus Thrombosis after "COVID-19 Vaccine AstraZeneca" Exposure. JCM 10:1599.

145. Wise J. 2021. Covid-19: European countries suspend use of Oxford-AstraZeneca vaccine after reports of blood clots. BMJ n699.

146. Mahase E. 2021. Covid-19: AstraZeneca vaccine is not linked to increased risk of blood clots, finds European Medicine Agency. BMJ n774.

147. Mahase E. 2021. Covid-19: US suspends Johnson and Johnson vaccine rollout over blood clots. BMJ n970.

148. Tanne JH. 2021. Covid-19: US authorises Johnson and Johnson vaccine again, ending pause in rollout. BMJ n1079.

149. Mahase E. 2021. Covid-19: Unusual blood clots are "very rare side effect" of Janssen vaccine, says EMA. BMJ n1046.

150. Oliver SE, Wallace M, See I, Mbaeyi S, Godfrey M, Hadler SC, Jatlaoui TC, Twentyman E, Hughes MM, Rao AK, Fiore A, Su JR, Broder KR, Shimabukuro T, Lale A, Shay DK, Markowitz LE, Wharton M, Bell BP, Brooks O, McNally V, Lee GM, Talbot HK, Daley MF. 2022. Use of the Janssen (Johnson & Johnson) COVID-19 Vaccine: Updated Interim Recommendations from the Advisory Committee on Immunization Practices — United States, December 2021. MMWR Morb Mortal Wkly Rep 71:90–95.

151. Shay DK, Gee J, Su JR, Myers TR, Marquez P, Liu R, Zhang B, Licata C, Clark TA, Shimabukuro TT. 2021. Safety Monitoring of the Janssen (Johnson & Johnson) COVID-19 Vaccine — United States, March–April 2021. MMWR Morb Mortal Wkly Rep 70:680–684.

152. Rosenblum HG, Hadler SC, Moulia D, Shimabukuro TT, Su JR, Tepper NK, Ess KC, Woo EJ, Mba-Jonas A, Alimchandani M, Nair N, Klein NP, Hanson KE, Markowitz LE, Wharton M, McNally VV, Romero JR, Talbot HK, Lee GM, Daley MF, Mbaeyi SA, Oliver SE. 2021. Use of COVID-19 Vaccines After Reports of Adverse Events Among Adult Recipients of Janssen (Johnson & Johnson) and mRNA COVID-19 Vaccines (Pfizer-BioNTech and Moderna): Update from the Advisory Committee on Immunization Practices — United States, July 2021. MMWR Morb Mortal Wkly Rep 70:1094–1099.

153. Pottegård A, Lund LC, Karlstad Ø, Dahl J, Andersen M, Hallas J, Lidegaard Ø, Tapia G, Gulseth HL, Ruiz PL-D, Watle SV, Mikkelsen AP, Pedersen L, Sørensen HT, Thomsen RW, Hviid A. 2021. Arterial events, venous thromboembolism, thrombocytopenia, and bleeding after vaccination with Oxford-AstraZeneca ChAdOx1-S in Denmark and Norway: population based cohort study. BMJ n1114.

154. Chan B, Odutayo A, Juni P, Stall NM, Bobos P, Brown AD, Grill A, Ivers N, Maltsev A, McGeer A, Miller KJ, Niel U, Razak F, Sander B, Sholzberg M, Slutsky AS, Morris AM, Pai M. 2021. Risk of Vaccine-Induced Thrombotic Thrombocytopenia (VITT) following the AstraZeneca/COVISHIELD Adenovirus Vector COVID-19 Vaccines. Ontario COVID-19 Science Advisory Table.

155. Baker AT, Boyd RJ, Sarkar D, Teijeira-Crespo A, Chan CK, Bates E, Waraich K, Vant J, Wilson E, Truong CD, Lipka-Lloyd M, Fromme P, Vermaas J, Williams D, Machiesky L, Heurich M, Nagalo BM, Coughlan L, Umlauf S, Chiu P-L, Rizkallah PJ, Cohen TS, Parker AL, Singharoy A, Borad MJ. 2021. ChAdOx1 interacts with CAR and PF4 with implications for thrombosis with thrombocytopenia syndrome. Sci Adv 7.

156. Schultz NH, Sørvoll IH, Michelsen AE, Munthe LA, Lund-Johansen F, Ahlen MT, Wiedmann M, Aamodt A-H, Skattør TH, Tjønnfjord GE, Holme PA. 2021. Thrombosis and Thrombocytopenia after ChAdOx1 nCoV-19 Vaccination. N Engl J Med 384:2124–2130.

157. 2021. EMA recommends COVID-19 Vaccine AstraZeneca for authorisation in the EU. https://www.ema.europa.eu/en/news/ema-recommends-covid-19-vaccine-astrazeneca-authorisation-eu.

158. Miesbach W, Makris M. 2020. COVID-19: Coagulopathy, Risk of Thrombosis, and the Rationale for Anticoagulation. Clin Appl Thromb Hemost 26:107602962093814.

159. Verbeke R, Lentacker I, De Smedt SC, Dewitte H. 2019. Three decades of messenger RNA vaccine development. Nano Today 28:100766.

160. Schlake T, Thess A, Fotin-Mleczek M, Kallen K-J. 2012. Developing mRNA-vaccine technologies. RNA Biology 9:1319–1330.

161. Martinon F, Krishnan S, Lenzen G, Magné R, Gomard E, Guillet J-G, Lévy J-P, Meulien P. 1993. Induction of virus-specific cytotoxic T lymphocytesin vivo by liposome-entrapped mRNA. European Journal of Immunology 23:1719–1722.

162. Zhang C, Maruggi G, Shan H, Li J. 2019. Advances in mRNA Vaccines for Infectious Diseases. Frontiers in Immunology 10:594.

163. Reichmuth AM, Oberli MA, Jaklenec A, Langer R, Blankschtein D. 2016. mRNA vaccine delivery using lipid nanoparticles. Therapeutic Delivery 7:319–334.

164. Iavarone C, O'hagan DT, Yu D, Delahaye NF, Ulmer JB. 2017. Mechanism of action of mRNA-based vaccines. Expert Review of Vaccines 16:871–881.

165. RNA vaccines: an introduction. PHG Foundation. https://www.phgfoundation.org/briefing/rna-vaccines. Retrieved 8 February 2021.

166. Crotty S. 2014. T Follicular Helper Cell Differentiation, Function, and Roles in Disease. Immunity 41:529–542.

167. Pardi N, Hogan MJ, Porter FW, Weissman D. 2018. mRNA vaccines — a new era in vaccinology. Nature Reviews Drug Discovery 17:261–279.

168. Stuart LM. 2021. In Gratitude for mRNA Vaccines. N Engl J Med 385:1436–1438.

169. Pardi N, Hogan MJ, Weissman D. 2020. Recent advances in mRNA vaccine technology. Current Opinion in Immunology 65:14–20.

170. Amanat F, Krammer F. 2020. SARS-CoV-2 Vaccines: Status Report. Immunity 52:583–589.

171.     Fonteilles-Drabek S, Reddy D, Wells TNC. 2017. Managing intellectual property to develop medicines for the world's poorest. Nat Rev Drug Discov 16:223–224.

172.     Study in Healthy Adults to Evaluate Gene Activation After Vaccination With GlaxoSmithKline (GSK) Biologicals' Candidate Tuberculosis (TB) Vaccine GSK 692342 - Full Text View - ClinicalTrials.gov. https://clinicaltrials.gov/ct2/show/NCT01669096. Retrieved 8 February 2021.

173.     Pardi N, Parkhouse K, Kirkpatrick E, McMahon M, Zost SJ, Mui BL, Tam YK, Karikó K, Barbosa CJ, Madden TD, Hope MJ, Krammer F, Hensley SE, Weissman D. 2018. Nucleoside-modified mRNA immunization elicits influenza virus hemagglutinin stalk-specific antibodies. Nature Communications 9:3361.

174.     Veiga N, Goldsmith M, Granot Y, Rosenblum D, Dammes N, Kedmi R, Ramishetti S, Peer D. 2018. Cell specific delivery of modified mRNA expressing therapeutic proteins to leukocytes. Nature Communications 9:4493.

175.     Vabret N, Britton GJ, Gruber C, Hegde S, Kim J, Kuksin M, Levantovsky R, Malle L, Moreira A, Park MD, Pia L, Risson E, Saffern M, Salomé B, Esai Selvan M, Spindler MP, Tan J, van der Heide V, Gregory JK, Alexandropoulos K, Bhardwaj N, Brown BD, Greenbaum B, Gümüş ZH, Homann D, Horowitz A, Kamphorst AO, Curotto de Lafaille MA, Mehandru S, Merad M, Samstein RM, Agrawal M, Aleynick M, Belabed M, Brown M, Casanova-Acebes M, Catalan J, Centa M, Charap A, Chan A, Chen ST, Chung J, Bozkus CC, Cody E, Cossarini F, Dalla E, Fernandez N, Grout J, Ruan DF, Hamon P, Humblin E, Jha D, Kodysh J, Leader A, Lin M, Lindblad K, Lozano-Ojalvo D, Lubitz G, Magen A, Mahmood Z, Martinez-Delgado G, Mateus-Tique J, Meritt E, Moon C, Noel J, O'Donnell T, Ota M, Plitt T, Pothula V, Redes J, Reyes Torres I, Roberto M, Sanchez-Paulete AR, Shang J, Schanoski AS, Suprun M, Tran M, Vaninov N, Wilk CM, Aguirre-Ghiso J, Bogunovic D, Cho J, Faith J, Grasset E, Heeger P, Kenigsberg E, Krammer F, Laserson U. 2020. Immunology of COVID-19: Current State of the Science. Immunity 52:910–941.

176.     Chien KR, Zangi L, Lui KO. 2014. Synthetic Chemically Modified mRNA (modRNA): Toward a New Technology Platform for Cardiovascular Biology and Medicine. Cold Spring Harbor Perspectives in Medicine 5:a014035–a014035.

177.     Pfizer and BioNTech Announce Early Positive Data from an Ongoing Phase 1/2 study of mRNA-based Vaccine Candidate Against SARS-CoV-2 | Pfizer. https://www.pfizer.com/news/press-release/press-release-detail/pfizer-and-biontech-announce-early-positive-data-ongoing-0. Retrieved 8 February 2021.

178.     National Institute of Allergy and Infectious Diseases (NIAID). 2020. Phase I, Open-Label, Dose-Ranging Study of the Safety and Immunogenicity of 2019-nCoV Vaccine (mRNA-1273) in Healthy Adults. NCT04283461. Clinical trial registration. clinicaltrials.gov.

179.     Funk CD, Laferrière C, Ardakani A. 2020. A Snapshot of the Global Race for Vaccines Targeting SARS-CoV-2 and the COVID-19 Pandemic. Frontiers in Pharmacology 11:937.

180. Ledford H. 2021. What the Moderna–NIH COVID vaccine patent fight means for research. Nature 600:200–201.

181. Diamond D. 2021. Moderna halts patent fight over coronavirus vaccine with federal government. Washington Post.

182. Goodman AS Brenda. 2022. Moderna files patent infringement lawsuits against Pfizer and BioNTech over mRNA Covid-19 vaccines. CNN. https://www.cnn.com/2022/08/26/health/moderna-pfizer-mrna-patent-lawsuit/index.html. Retrieved 5 December 2022.

183. Polack FP, Thomas SJ, Kitchin N, Absalon J, Gurtman A, Lockhart S, Perez JL, Pérez Marc G, Moreira ED, Zerbini C, Bailey R, Swanson KA, Roychoudhury S, Koury K, Li P, Kalina WV, Cooper D, Frenck RW, Hammitt LL, Türeci Ö, Nell H, Schaefer A, Ünal S, Tresnan DB, Mather S, Dormitzer PR, Şahin U, Jansen KU, Gruber WC. 2020. Safety and Efficacy of the BNT162b2 mRNA Covid-19 Vaccine. New England Journal of Medicine 383:2603–2615.

184. Baden LR, El Sahly HM, Essink B, Kotloff K, Frey S, Novak R, Diemert D, Spector SA, Rouphael N, Creech CB, McGettigan J, Khetan S, Segall N, Solis J, Brosz A, Fierro C, Schwartz H, Neuzil K, Corey L, Gilbert P, Janes H, Follmann D, Marovich M, Mascola J, Polakowski L, Ledgerwood J, Graham BS, Bennett H, Pajon R, Knightly C, Leav B, Deng W, Zhou H, Han S, Ivarsson M, Miller J, Zaks T. 2020. Efficacy and Safety of the mRNA-1273 SARS-CoV-2 Vaccine. New England Journal of Medicine NEJMoa2035389.

185. Commissioner O of the. 2020. FDA Takes Key Action in Fight Against COVID-19 By Issuing Emergency Use Authorization for First COVID-19 Vaccine. FDA. https://www.fda.gov/news-events/press-announcements/fda-takes-key-action-fight-against-covid-19-issuing-emergency-use-authorization-first-covid-19. Retrieved 8 February 2021.

186. Oliver SE. 2021. The Advisory Committee on Immunization Practices' Interim Recommendation for Use of Moderna COVID-19 Vaccine — United States, December 2020. MMWR Morbidity and Mortality Weekly Report 69.

187. Zimmer C, Corum J, Wee S-L, Kristoffersen M. 2020. Coronavirus Vaccine Tracker. The New York Times.

188. Thompson MG, Burgess JL, Naleway AL, Tyner H, Yoon SK, Meece J, Olsho LEW, Caban-Martinez AJ, Fowlkes AL, Lutrick K, Groom HC, Dunnigan K, Odean MJ, Hegmann K, Stefanski E, Edwards LJ, Schaefer-Solle N, Grant L, Ellingson K, Kuntz JL, Zunie T, Thiese MS, Ivacic L, Wesley MG, Mayo Lamberte J, Sun X, Smith ME, Phillips AL, Groover KD, Yoo YM, Gerald J, Brown RT, Herring MK, Joseph G, Beitel S, Morrill TC, Mak J, Rivers P, Poe BP, Lynch B, Zhou Y, Zhang J, Kelleher A, Li Y, Dickerson M, Hanson E, Guenther K, Tong S, Bateman A, Reisdorf E, Barnes J, Azziz-Baumgartner E, Hunt DR, Arvay ML, Kutty P, Fry AM, Gaglani M. 2021. Prevention and Attenuation of Covid-19 with the BNT162b2 and mRNA-1273 Vaccines. N Engl J Med 385:320–329.

189.     2021. The effects of virus variants on COVID-19 vaccines. https://www.who.int/news-room/feature-stories/detail/the-effects-of-virus-variants-on-covid-19-vaccines. Retrieved 5 December 2022.

190.     Puranik A, Lenehan PJ, Silvert E, Niesen MJM, Corchado-Garcia J, O'Horo JC, Virk A, Swift MD, Halamka J, Badley AD, Venkatakrishnan AJ, Soundararajan V. 2021. Comparison of two highly-effective mRNA vaccines for COVID-19 during periods of Alpha and Delta variant prevalence. Cold Spring Harbor Laboratory.

191.     Tseng HF, Ackerson BK, Luo Y, Sy LS, Talarico CA, Tian Y, Bruxvoort KJ, Tubert JE, Florea A, Ku JH, Lee GS, Choi SK, Takhar HS, Aragones M, Qian L. 2022. Effectiveness of mRNA-1273 against SARS-CoV-2 Omicron and Delta variants. Nat Med 28:1063–1071.

192.     Collie S, Champion J, Moultrie H, Bekker L-G, Gray G. 2022. Effectiveness of BNT162b2 Vaccine against Omicron Variant in South Africa. N Engl J Med 386:494–496.

193.     Andrews N, Stowe J, Kirsebom F, Toffa S, Rickeard T, Gallagher E, Gower C, Kall M, Groves N, O'Connell A-M, Simons D, Blomquist PB, Zaidi A, Nash S, Iwani Binti Abdul Aziz N, Thelwall S, Dabrera G, Myers R, Amirthalingam G, Gharbia S, Barrett JC, Elson R, Ladhani SN, Ferguson N, Zambon M, Campbell CNJ, Brown K, Hopkins S, Chand M, Ramsay M, Lopez Bernal J. 2022. Covid-19 Vaccine Effectiveness against the Omicron (B.1.1.529) Variant. N Engl J Med 386:1532–1546.

194.     Coronavirus in the U.S.: Latest Map and Case Count. The New York Times. https://www.nytimes.com/interactive/2021/us/covid-cases.html. Retrieved 11 March 2022.

195.     Tan S. Analysis | Four charts that analyze how omicron's wave compares to previous coronavirus peaks. Washington Post. https://www.washingtonpost.com/health/interactive/2022/omicron-comparison-cases-deaths-hospitalizations/. Retrieved 5 December 2022.

196.     Abu Mouch S, Roguin A, Hellou E, Ishai A, Shoshan U, Mahamid L, Zoabi M, Aisman M, Goldschmid N, Berar Yanay N. 2021. Myocarditis following COVID-19 mRNA vaccination. Vaccine 39:3790–3793.

197.     Kim HW, Jenista ER, Wendell DC, Azevedo CF, Campbell MJ, Darty SN, Parker MA, Kim RJ. 2021. Patients With Acute Myocarditis Following mRNA COVID-19 Vaccination. JAMA Cardiol 6:1196.

198.     Mevorach D, Anis E, Cedar N, Bromberg M, Haas EJ, Nadir E, Olsha-Castell S, Arad D, Hasin T, Levi N, Asleh R, Amir O, Meir K, Cohen D, Dichtiar R, Novick D, Hershkovitz Y, Dagan R, Leitersdorf I, Ben-Ami R, Miskin I, Saliba W, Muhsen K, Levi Y, Green MS, Keinan-Boker L, Alroy-Preis S. 2021. Myocarditis after BNT162b2 mRNA Vaccine against Covid-19 in Israel. N Engl J Med 385:2140–2149.

199.     Goddard K, Hanson KE, Lewis N, Weintraub E, Fireman B, Klein NP. 2022. Incidence of Myocarditis/Pericarditis Following mRNA COVID-19 Vaccination Among Children and Younger Adults in the United States. Ann Intern Med https://doi.org/10.7326/m22-2274.

200. Matta A, Kunadharaju R, Osman M, Jesme C, McMiller Z, Johnson EM, Matta D, Kallamadi R, Bande D. 2021. Clinical Presentation and Outcomes of Myocarditis Post mRNA Vaccination: A Meta-Analysis and Systematic Review. Cureus https://doi.org/10.7759/cureus.19240.

201. Kim JH, Marks F, Clemens JD. 2021. Looking beyond COVID-19 vaccine phase 3 trials. Nat Med 27:205–211.

202. Commissioner O of the. 2022. Coronavirus (COVID-19) Update: FDA Recommends Inclusion of Omicron BA.4/5 Component for COVID-19 Vaccine Booster Doses. FDA. https://www.fda.gov/news-events/press-announcements/coronavirus-covid-19-update-fda-recommends-inclusion-omicron-ba45-component-covid-19-vaccine-booster. Retrieved 5 December 2022.

203. 2022. FDA Panel Gives Thumbs Up to Omicron-Containing COVID Boosters. https://www.medpagetoday.com/infectiousdisease/covid19vaccine/99493. Retrieved 5 December 2022.

204. 2022. Fall 2022 COVID-19 Vaccine Strain Composition Selection Recommendation. US Food and Drug Administration. https://www.fda.gov/media/159597/download.

205. Erman M. 2022. U.S. FDA to use existing Omicron booster data to review shots targeting new subvariants -official. Reuters.

206. Tseng HF, Ackerson BK, Bruxvoort KJ, Sy LS, Tubert JE, Lee GS, Ku JH, Florea A, Luo Y, Qiu S, Choi SK, Takhar HS, Aragones M, Paila YD, Chavers S, Talarico CA, Qian L. 2022. Effectiveness of mRNA-1273 against infection and COVID-19 hospitalization with SARS-CoV-2 Omicron subvariants: BA.1, BA.2, BA.2.12.1, BA.4, and BA.5. Cold Spring Harbor Laboratory.

207. Hardt K, Vandebosch A, Sadoff J, Gars ML, Truyers C, Lowson D, Van Dromme I, Vingerhoets J, Kamphuis T, Scheper G, Ruiz-Guiñazú J, Faust SN, Spinner CD, Schuitemaker H, Van Hoof J, Douoguih M, Struyf F. 2022. Efficacy and Safety of a Booster Regimen of Ad26.COV2.S Vaccine against Covid-19. Cold Spring Harbor Laboratory.

208. Dolzhikova I, Iliukhina A, Kovyrshina A, Kuzina A, Gushchin V, Siniavin A, Pochtovyi A, Shidlovskaya E, Kuznetsova N, Megeryan M, Dzharullaeva A, Erokhova A, Izhaeva F, Grousova D, Botikov A, Shcheblyakov D, Tukhvatulin A, Zubkova O, Logunov D, Gintsburg A. 2021. Sputnik Light booster after Sputnik V vaccination induces robust neutralizing antibody response to B.1.1.529 (Omicron) SARS-CoV-2 variant. Cold Spring Harbor Laboratory.

209. Flaxman A, Marchevsky NG, Jenkin D, Aboagye J, Aley PK, Angus B, Belij-Rammerstorfer S, Bibi S, Bittaye M, Cappuccini F, Cicconi P, Clutterbuck EA, Davies S, Dejnirattisai W, Dold C, Ewer KJ, Folegatti PM, Fowler J, Hill AVS, Kerridge S, Minassian AM, Mongkolsapaya J, Mujadidi YF, Plested E, Ramasamy MN, Robinson H, Sanders H, Sheehan E, Smith H, Snape MD, Song R, Woods D, Screaton G, Gilbert SC, Voysey M, Pollard AJ, Lambe T, Adlou S, Aley R, Ali A, Anslow R, Baker M, Baker P, Barrett JR, Bates L, Beadon K, Beckley R, Bell J, Bellamy D, Beveridge A, Bissett C, Blackwell L, Bletchly H, Boyd A, Bridges-Webb A, Brown C, Byard N, Camara S, Cifuentes Gutierrez L, Collins AM, Cooper R, Crocker WEM, Darton TC, Davies H, Davies J, Demissie T, Di Maso C, Dinesh T, Donnellan FR, Douglas AD, Drake-Brockman R, Duncan CJA, Elias SC, Emary KRW, Ghulam Farooq M, Faust SN, Felle S, Ferreira D, Ferreira Da Silva C, Finn A, Ford KJ, Francis E, Furze J, Fuskova M, Galiza E, Gibertoni Cruz A, Godfrey L, Goodman AL, Green C, Green CA, Greenwood N, Harrison D, Hart TC, Hawkins S, Heath PT, Hill H, Hillson K, Horsington B, Hou MM, Howe E, Howell N, Joe C, Jones E, Kasanyinga M, Keen J, Kelly S, Kerr D, Khan L, Khozoee B, Kinch J, Kinch P, Koleva S, Kwok J, Larkworthy CW, Lawrie AM, Lazarus R, Lees EA, Li G, Libri V, Lillie PJ, Linder A, Long F, Lopez Ramon R, Mabbett R, Makinson R, Marinou S, Marlow E, Marshall JL, Mazur O, McEwan J, McGregor AC, Mokaya J, Morey E, Morshead G, Morter R, Muller J, Mweu P, Noristani R, Owino N, Polo Peralta Alvarez M, Platt A, Pollock KM, Poulton I, Provstgaard-Morys S, Pulido-Gomez D, Rajan M, Ramos Lopez F, Ritchie A, Roberts H, Rollier C, Rudiansyah I, Sanders K, Saunders JE, Seddiqi S, Sharpe HR, Shaw R, Silva-Reyes L, Singh N, Smith DJ, Smith CC, Smith A, Spencer AJ, Stuart ASV, Sutherland R, Szigeti A, Tang K, Thomas M, Thomas TM, Thompson A, Thomson EC, Török EM, Toshner M, Tran N, Trivett R, Turnbull I, Turner C, Turner DPJ, Ulaszewska M, Vichos I, Walker L, Watson ME, Whelan C, White R, Williams SJ, Williams CJA, Wright D, Yao A. 2021. Reactogenicity and immunogenicity after a late second dose or a third dose of ChAdOx1 nCoV-19 in the UK: a substudy of two randomised controlled trials (COV001 and COV002). The Lancet 398:981–990.

210. Regev-Yochay G, Gonen T, Gilboa M, Mandelboim M, Indenbaum V, Amit S, Meltzer L, Asraf K, Cohen C, Fluss R, Biber A, Nemet I, Kliker L, Joseph G, Doolman R, Mendelson E, Freedman LS, Harats D, Kreiss Y, Lustig Y. 2022. 4th Dose COVID mRNA Vaccines' Immunogenicity & Efficacy Against Omicron VOC. Cold Spring Harbor Laboratory.

211. Chiu N-C, Chi H, Tu Y-K, Huang Y-N, Tai Y-L, Weng S-L, Chang L, Huang DT-N, Huang F-Y, Lin C-Y. 2021. To mix or not to mix? A rapid systematic review of heterologous prime–boost covid-19 vaccination. Expert Review of Vaccines 20:1211–1220.

212. Atmar RL, Lyke KE, Deming ME, Jackson LA, Branche AR, El Sahly HM, Rostad CA, Martin JM, Johnston C, Rupp RE, Mulligan MJ, Brady RC, Frenck RW Jr., Bäcker M, Kottkamp AC, Babu TM, Rajakumar K, Edupuganti S, Dobrzynski D, Coler RN, Posavad CM, Archer JI, Crandon S, Nayak SU, Szydlo D, Zemanek JA, Dominguez Islas CP, Brown ER, Suthar MS, McElrath MJ, McDermott AB, O'Connell SE, Montefiori DC, Eaton A, Neuzil KM, Stephens DS, Roberts PC, Beigel JH. 2022. Homologous and Heterologous Covid-19 Booster Vaccinations. N Engl J Med 386:1046–1057.

213. Jara A, Undurraga EA, Zubizarreta JR, González C, Pizarro A, Acevedo J, Leo K, Paredes F, Bralic T, Vergara V, Mosso M, Leon F, Parot I, Leighton P, Suárez P, Rios JC, García-Escorza H, Araos R. 2022. Effectiveness of homologous and heterologous booster doses for an inactivated SARS-CoV-2 vaccine: a large-scale prospective cohort study. The Lancet Global Health 10:e798–e806.

214. Sapkota B, Saud B, Shrestha R, Al-Fahad D, Sah R, Shrestha S, Rodriguez-Morales AJ. 2021. Heterologous prime–boost strategies for COVID-19 vaccines. Journal of Travel Medicine https://doi.org/10.1093/jtm/taab191.

215.	Assawakosri S, Kanokudom S, Suntronwong N, Auphimai C, Nilyanimit P, Vichaiwattana P, Thongmee T, Duangchinda T, Chantima W, Pakchotanon P, Srimuan D, Thatsanatorn T, Klinfueng S, Yorsaeng R, Sudhinaraset N, Wanlapakorn N, Mongkolsapaya J, Honsawek S, Poovorawan Y. 2022. Neutralizing Activities Against the Omicron Variant After a Heterologous Booster in Healthy Adults Receiving Two Doses of CoronaVac Vaccination. The Journal of Infectious Diseases 226:1372–1381.

216.	Accorsi EK, Britton A, Shang N, Fleming-Dutra KE, Link-Gelles R, Smith ZR, Derado G, Miller J, Schrag SJ, Verani JR. 2022. Effectiveness of Homologous and Heterologous Covid-19 Boosters against Omicron. N Engl J Med 386:2433–2435.

217.	Munro APS, Janani L, Cornelius V, Aley PK, Babbage G, Baxter D, Bula M, Cathie K, Chatterjee K, Dodd K, Enever Y, Gokani K, Goodman AL, Green CA, Harndahl L, Haughney J, Hicks A, van der Klaauw AA, Kwok J, Lambe T, Libri V, Llewelyn MJ, McGregor AC, Minassian AM, Moore P, Mughal M, Mujadidi YF, Murira J, Osanlou O, Osanlou R, Owens DR, Pacurar M, Palfreeman A, Pan D, Rampling T, Regan K, Saich S, Salkeld J, Saralaya D, Sharma S, Sheridan R, Sturdy A, Thomson EC, Todd S, Twelves C, Read RC, Charlton S, Hallis B, Ramsay M, Andrews N, Nguyen-Van-Tam JS, Snape MD, Liu X, Faust SN, Riordan A, Ustianowski A, Rogers CA, Hughes S, Longshaw L, Stockport J, Hughes R, Grundy L, Tudor Jones L, Guha A, Snashall E, Eadsforth T, Reeder S, Storton K, Munusamy M, Tandy B, Egbo A, Cox S, Ahmed NN, Shenoy A, Bousfield R, Wixted D, Gutteridge H, Mansfield B, Herbert C, Holliday K, Calderwood J, Barker D, Brandon J, Tulloch H, Colquhoun S, Thorp H, Radford H, Evans J, Baker H, Thorpe J, Batham S, Hailstone J, Phillips R, Kumar D, Westwell F, Makia F, Hopkins N, Barcella L, Mpelembue M, dabagh M, lang M, khan F, Adebambo O, Chita S, Corrah T, Whittington A, John L, Roche S, Wagstaff L, Farrier A, Bisnauthsing K, Serafimova T, Nanino E, Cooney E, Wilson-Goldsmith J, Nguyen H, Mazzella A, Jackson B, Aslam S, Bawa T, Broadhead S, Farooqi S, Piper J, Weighell R, Pickup L, Shamtally D, Domingo J, Kourampa E, Hale C, Gibney J, Stackpoole M, Rashid-Gardner Z, Lyon R, McDonnell C, Cole C, Stewart A, McMillan G, Savage M, Beckett H, Moorbey C, Desai A, Brown C, Naker K, Qureshi E, Trinham C, Sabine C, Moore S, Hurdover S, Justice E, Smith D, Plested E, Ferreira Da Silva C, White R, Robinson H, Cifuentes L, Morshead G, Drake-Brockman R, Kinch P, Kasanyinga M, Clutterbuck EA, Bibi S, Stuart AS, Shaw RH, Singh M, Champaneri T, Irwin M, Khan M, Kownacka A, Nabunjo M, Osuji C, Hladkiwskyj J, Galvin D, Patel G, Mouland J, Longhurst B, Moon M, Giddins B, Pereira Dias Alves C, Richmond L, Minnis C, Baryschpolec S, Elliott S, Fox L, Graham V, Baker N, Godwin K, Buttigieg K, Knight C, Brown P, Lall P, Shaik I, Chiplin E, Brunt E, Leung S, Allen L, Thomas S, Fraser S, Choi B, Gouriet J, Freedman A, Perkins J, Gowland A, Macdonald J, Seenan JP, Starinskij I, Seaton A, Peters E, Singh S, Gardside B, Bonnaud A, Davies C, Gordon E, Keenan S, Hall J, Wilkins S, Tasker S, James R, Seath I, Littlewood K, Newman J, Boubriak I, Suggitt D, Haydock H, Bennett S, Woodyatt W, Hughes K, Bell J, Coughlan T, van Welsenes D, Kamal M, Cooper C, Tunstall S, Ronan N, Cutts R, Dare T, Yim YTN, Whittley S, Ricamara M, Hamal S, Adams K, Baker H, Driver K, Turner N, Rawlins T, Roy S, Merida-Morillas M, Sakagami Y, Andrews A, Goncalves cordeiro L, Stokes M, Ambihapathy W, Spencer J, Parungao N, Berry L, Cullinane J, Presland L, Ross-Russell A, Warren S, Baker J, Oliver A, Buadi A, Lee K, Haskell L, Romani R, Bentley I, Whitbred T, Fowler S, Gavin J, Magee A, Watson T, Nightingale K, Marius P, Summerton E, Locke E, Honey T, Lingwood A, de la Haye A, Elliott RS, Underwood K, King M, Davies-Dear S, Horsfall E, Chalwin O, Burton H,

Edwards CJ, Welham B, Garrahy S, Hall F, Ladikou E, Mullan D, Hansen D, Campbell M, Dos Santos F, Habash-Bailey H, Lakeman N, Branney D, Vamplew L, Hogan A, Frankham J, Wiselka M, Vail D, Wenn V, Renals V, Ellis K, Lewis-Taylor J, Magan J, Hardy A, Appleby K. 2021. Safety and immunogenicity of seven COVID-19 vaccines as a third dose (booster) following two doses of ChAdOx1 nCov-19 or BNT162b2 in the UK (COV-BOOST): a blinded, multicentre, randomised, controlled, phase 2 trial. The Lancet 398:2258–2276.

218.    2021. Overview of EU/EEA country recommendations on COVID-19 vaccination with Vaxzevria, and a scoping review of evidence to guide decision-making. European Centre for Disease Prevention and Control. https://www.ecdc.europa.eu/en/publications-data/overview-eueea-country-recommendations-covid-19-vaccination-vaxzevria-and-scoping. Retrieved 5 December 2022.

219.    Duarte-Salles T, Prieto-Alhambra D. 2021. Heterologous vaccine regimens against COVID-19. The Lancet 398:94–95.

220.    Moderna Announces Omicron-Containing Bivalent Booster Candidate mRNA-1273.214 Demonstrates Superior Antibody Response Against Omicron. https://investors.modernatx.com/news/news-details/2022/Moderna-Announces-Omicron-Containing-Bivalent-Booster-Candidate-mRNA-1273.214-Demonstrates-Superior-Antibody-Response-Against-Omicron/default.aspx. Retrieved 5 December 2022.

221.    Choi A, Koch M, Wu K, Chu L, Ma L, Hill A, Nunna N, Huang W, Oestreicher J, Colpitts T, Bennett H, Legault H, Paila Y, Nestorova B, Ding B, Montefiori D, Pajon R, Miller JM, Leav B, Carfi A, McPhee R, Edwards DK. 2021. Safety and immunogenicity of SARS-CoV-2 variant mRNA vaccine boosters in healthy adults: an interim analysis. Nat Med 27:2025–2031.

222.    Pajon R, Doria-Rose NA, Shen X, Schmidt SD, O'Dell S, McDanal C, Feng W, Tong J, Eaton A, Maglinao M, Tang H, Manning KE, Edara V-V, Lai L, Ellis M, Moore KM, Floyd K, Foster SL, Posavad CM, Atmar RL, Lyke KE, Zhou T, Wang L, Zhang Y, Gaudinski MR, Black WP, Gordon I, Guech M, Ledgerwood JE, Misasi JN, Widge A, Sullivan NJ, Roberts PC, Beigel JH, Korber B, Baden LR, El Sahly H, Chalkias S, Zhou H, Feng J, Girard B, Das R, Aunins A, Edwards DK, Suthar MS, Mascola JR, Montefiori DC. 2022. SARS-CoV-2 Omicron Variant Neutralization after mRNA-1273 Booster Vaccination. N Engl J Med 386:1088–1091.

223.    Pfizer and BioNTech Announce Omicron-Adapted COVID-19 Vaccine Candidates Demonstrate High Immune Response Against Omicron | Pfizer. https://www.pfizer.com/news/press-release/press-release-detail/pfizer-and-biontech-announce-omicron-adapted-covid-19. Retrieved 5 December 2022.

224.    Moxon R, Reche PA, Rappuoli R. 2019. Editorial: Reverse Vaccinology. Front Immunol 10.

225.    Kudchodkar SB, Choi H, Reuschel EL, Esquivel R, Jin-Ah Kwon J, Jeong M, Maslow JN, Reed CC, White S, Kim JJ, Kobinger GP, Tebas P, Weiner DB, Muthumani K. 2018. Rapid response to an emerging infectious disease – Lessons learned from development of a synthetic DNA vaccine targeting Zika virus. Microbes and Infection 20:676–684.

226.	Olena A. 2020. Newer Vaccine Technologies Deployed to Develop COVID-19 Shot. The Scientist Magazine. https://www.the-scientist.com/news-opinion/newer-vaccine-technologies-deployed-to-develop-covid-19-shot-67152.

227.	Pulliam JRC, van Schalkwyk C, Govender N, Gottberg A von, Cohen C, Groome MJ, Dushoff J, Mlisana K, Moultrie H. 2021. Increased risk of SARS-CoV-2 reinfection associated with emergence of Omicron in South Africa. Cold Spring Harbor Laboratory.

228.	2022-07-15 12:20 | Archive of CDC Covid Pages. https://public4.pagefreezer.com/browse/CDC%20Covid%20Pages/15-07-2022T12:20/https://www.cdc.gov/coronavirus/2019-ncov/science/science-briefs/scientific-brief-omicron-variant.html. Retrieved 5 December 2022.

229.	Planas D, Bruel T, Grzelak L, Guivel-Benhassine F, Staropoli I, Porrot F, Planchais C, Buchrieser J, Rajah MM, Bishop E, Albert M, Donati F, Prot M, Behillil S, Enouf V, Maquart M, Smati-Lafarge M, Varon E, Schortgen F, Yahyaoui L, Gonzalez M, De Sèze J, Péré H, Veyer D, Sève A, Simon-Lorière E, Fafi-Kremer S, Stefic K, Mouquet H, Hocqueloux L, van der Werf S, Prazuck T, Schwartz O. 2021. Sensitivity of infectious SARS-CoV-2 B.1.1.7 and B.1.351 variants to neutralizing antibodies. Nat Med 27:917–924.

230.	Wang P, Nair MS, Liu L, Iketani S, Luo Y, Guo Y, Wang M, Yu J, Zhang B, Kwong PD, Graham BS, Mascola JR, Chang JY, Yin MT, Sobieszczyk M, Kyratsous CA, Shapiro L, Sheng Z, Huang Y, Ho DD. 2021. Antibody resistance of SARS-CoV-2 variants B.1.351 and B.1.1.7. Nature 593:130–135.

231.	Edara VV, Hudson WH, Xie X, Ahmed R, Suthar MS. 2021. Neutralizing Antibodies Against SARS-CoV-2 Variants After Infection and Vaccination. JAMA 325:1896.

232.	Classification of Omicron (B.1.1.529): SARS-CoV-2 Variant of Concern. https://www.who.int/news/item/26-11-2021-classification-of-omicron-(b.1.1.529)-sars-cov-2-variant-of-concern. Retrieved 5 December 2022.

233.	CDC. 2020. COVID Data Tracker. Centers for Disease Control and Prevention. https://covid.cdc.gov/covid-data-tracker. Retrieved 5 December 2022.

234.	Kuhlmann C, Mayer CK, Claassen M, Maponga TG, Sutherland AD, Suliman T, Shaw M, Preiser W. 2021. Breakthrough Infections with SARS-CoV-2 Omicron Variant Despite Booster Dose of mRNA Vaccine. SSRN Journal https://doi.org/10.2139/ssrn.3981711.

235.	Self WH, Tenforde MW, Rhoads JP, Gaglani M, Ginde AA, Douin DJ, Olson SM, Talbot HK, Casey JD, Mohr NM, Zepeski A, McNeal T, Ghamande S, Gibbs KW, Files DC, Hager DN, Shehu A, Prekker ME, Erickson HL, Gong MN, Mohamed A, Henning DJ, Steingrub JS, Peltan ID, Brown SM, Martin ET, Monto AS, Khan A, Hough CL, Busse LW, ten Lohuis CC, Duggal A, Wilson JG, Gordon AJ, Qadir N, Chang SY, Mallow C, Rivas C, Babcock HM, Kwon JH, Exline MC, Halasa N, Chappell JD, Lauring AS, Grijalva CG, Rice TW, Jones ID, Stubblefield WB, Baughman A, Womack KN, Lindsell CJ, Hart KW, Zhu Y, Mills L, Lester SN, Stumpf MM, Naioti EA, Kobayashi M, Verani JR, Thornburg NJ, Patel MM, Calhoun N, Murthy K, Herrick J,

McKillop A, Hoffman E, Zayed M, Smith M, Seattle N, Ettlinger J, Priest E, Thomas J, Arroliga A, Beeram M, Kindle R, Kozikowski L-A, De Souza L, Ouellette S, Thornton-Thompson S, Mehkri O, Ashok K, Gole S, King A, Poynter B, Stanley N, Hendrickson A, Maruggi E, Scharber T, Jorgensen J, Bowers R, King J, Aston V, Armbruster B, Rothman RE, Nair R, Chen J-TT, Karow S, Robart E, Maldonado PN, Khan M, So P, Levitt J, Perez C, Visweswaran A, Roque J, Rivera A, Angeles L, Frankel T, Angeles L, Goff J, Huynh D, Howell M, Friedel J, Tozier M, Driver C, Carricato M, Foster A, Nassar P, Stout L, Sibenaller Z, Walter A, Mares J, Olson L, Clinansmith B, Rivas C, Gershengorn H, McSpadden E, Truscon R, Kaniclides A, Thomas L, Bielak R, Valvano WD, Fong R, Fitzsimmons WJ, Blair C, Valesano AL, Gilbert J, Crider CD, Steinbock KA, Paulson TC, Anderson LA, Kampe C, Johnson J, McHenry R, Blair M, Conway D, LaRose M, Landreth L, Hicks M, Parks L, Bongu J, McDonald D, Cass C, Seiler S, Park D, Hink T, Wallace M, Burnham C-A, Arter OG. 2021. Comparative Effectiveness of Moderna, Pfizer-BioNTech, and Janssen (Johnson & Johnson) Vaccines in Preventing COVID-19 Hospitalizations Among Adults Without Immunocompromising Conditions — United States, March–August 2021. MMWR Morb Mortal Wkly Rep 70:1337–1343.

236.    Mueller B, Robbins R. 2021. Where a Vast Global Vaccination Program Went Wrong. The New York Times.

237.    Sheridan C. 2021. Innovators target vaccines for variants and shortages in global South. Nat Biotechnol 39:393–396.

238.    Nohynek H, Wilder-Smith A. 2022. Does the World Still Need New Covid-19 Vaccines? N Engl J Med 386:2140–2142.